\documentstyle[epsf]{mn}
\def\epsfsize#1#2{\hsize}

\def\sgr{Sgr dSph }

\begin{document}

\title[The Sgr Dwarf Galaxy Survey II]
{The Sagittarius Dwarf Spheroidal Galaxy Survey (SDGS) II: 
The Stellar Content and  
Constraints on the Star Formation History.\thanks{Based on data 
taken at the New Technology Telescope - ESO, La Silla.}}

\author[M. Bellazzini et al.]
       {M. Bellazzini,$^1$
       F.R. Ferraro,$^1$
       and R. Buonanno,$^2$\\
        $^1$Osservatorio Astronomico di Bologna, Via Ranzani 1, 40127 
Bologna, ITALY\\
        $^2$Osservatorio Astronomico di Roma, Via dell'Osservatorio 2, 
00040, Monte Porzio Catone, Roma, ITALY}

\date{Accepted 
      Received ;
      in original form }
\pubyear{1998}

\maketitle

\begin{abstract}

A detailed study of the Star Formation History of the \sgr galaxy is performed
through the analysis of the data from the Sagittarius Dwarf Galaxy Survey
(SDGS; Bellazzini, Ferraro \& Buonanno 1999). Accurate
statistical decontamination of the SDGS Color - Magnitude diagrams allow us to
obtain many useful constraints on the age and metal content of the Sgr stellar
populations in three different region of the galaxy.

A coarse metallicity distribution of Sgr stars is derived, ranging from
$[Fe/H]\sim -2.0$ to $[Fe/H]\sim -0.7$, the upper limit being somewhat higher
in the central region of the galaxy. A qualitative global fit to all the
observed CMD features is attempted, and a general scheme for the Star Formation
History of the \sgr is derived. According to this scheme, star formation began
at very early time from a low metal content Inter Stellar Medium and lasted for
several Gyr, coupled with progressive chemical enrichment. The Star Formation
Rate (SFR) had a peak from 8 to 10 gyr ago when the mean metallicity was in the range
$-1.3\le [Fe/H] \le -0.7$. After that maximum, the SFR rapidly decreased and
very low rate star formation took place until $\sim 1-0.5$ Gyr ago.
 
\end{abstract}


\section{Introduction}

The recently discovered Sagittarius dwarf Spheroidal galaxy (Ibata
et al. 1994; hereafter IGI-I) is the nearest satellite galaxy of the Milky Way.
The structure of the \sgr appears to be strongly disturbed by the interaction
with the Galaxy and we are probably witnessing the ongoing process of merging
of a dwarf galactic sub-unit into a main giant galaxy 
[see Ibata et al. 1995 (IGI-II); Ibata et al. 1997 (IWGIS); Mateo et al. 1995a
(MUSKKK); Mateo et al. 1995b (MKSKKU); Mateo et al. 1996; 
Sarajedini \& Layden 1995 (SL95); Marconi et al. 1998 (MAL), 
Whitelock, Irwin \& Catchpole 1996 (WIC); Fahlman et al. 1998 (FAL); 
Montegriffo et al. 1998 (MoAL), for further discussions and references]. 
  
Thus Sgr represents a unique opportunity to study in detail the stellar content
and the Star Formation History (SFH) of a dwarf spheroidal galaxy, and the 
possible influence of the interactions with a parent galaxy on the SFH itself.
Furthermore the study of the Sgr galaxy can be an ideal testbed for many
relevant open problems in astrophysics, as, for instance, the comprehension of 
the evolutionary path of dwarf galaxies (Grebel 1998), the dark matter content
of dSph's and the formation of the Milky Way halo and/or bulge (Johnston 1998,
Fusi Pecci et al. 1995, Majewsky 1996).

However, the detailed analysis of the Sgr stellar content represent a
hard observational challenge, because of {\it (a)} the very low surface 
brightness (i.e. number of stars per unit area)
of the galaxy, and {\it (b)} the high degree of contamination by foreground
stars belonging to the Milky Way Bulge and Disc.  

In a companion paper (Bellazzini, Ferraro \& Buonanno 1999; hereafter PAP-I) we
presented the data and the first results of a large photometric survey, the
Sagittarius Dwarf Galaxy Survey (SDGS), aimed to the study of the stellar
content of the Sgr galaxy and specifically planned to attempt overcoming the
problems {\em (a)} and {\em (b)} described above.   

The scientific rationale of the SDGS is:

\begin{itemize}

\item To collect photometric data for a large number of stars in the direction
of Sgr, in order to sample adequately the Post Main Sequence (PMS) branches 
in the Color - Magnitude Diagram (CMD) and providing the natural
complement to the HST photometry by Mighell et al. (1997).  

\item To obtain a significant sample of the foreground population contaminating
the Sgr fields, in order to attempt to disentangle the CMD features of the Sgr
galaxy from those associated with the Galactic field.

\item To observe different regions of the Sgr galaxy searching for possible
spatial variations in the stellar content.

\item To provide constraints on the age and
metallicity of the Sgr stellar populations, trying to reconstruct the main
events of the SFH. 
    
\end{itemize}
 
The main characteristics of the SDGS and the results of PAP-I will be  
summarized in Sect. 2, in order to set the basis for the analysis performed in
the present paper, in which we discuss the decontaminated CMDs and their main
features.

In Sect. 3 the decontamination method is presented and
tested. Sect. 4 is devoted to the detailed analysis of the decontaminated CMDs,
and to a comparison between the stellar content of the different fields.
Sect. 5 reports the results about the metallicity distribution of Sgr stars, and
in Sect. 6 we report age estimates. A new general scheme for the Star Formation
History in the Sgr galaxy is proposed.  
Discussion and conclusions are presented in Sect. 7.

\section{Nomenclature and results of PAP-I}

In this section we briefly summarize the contents of PAP-I for an easier 
comprehension of the present paper. 

\subsection{Observed fields and completeness of the samples}

Three wide fields ($\sim 35\times 9 ~arcmin^2$) have been observed toward the 
Sgr galaxy:

\begin{itemize}

\item SGR12, located at $(l;b)\sim(6;-14)$, sampling the densest clump of the
Sgr galaxy, near the M54 globular cluster (see IGI-II and IWGIS). 
Calibrated V and I magnitude has been obtained for a total 
of 25793 stars, with a typical limiting magnitude of $V\sim 21.8$. Because of
bad weather conditions, for a considerable region of this field (nearly 2/5) 
we got acceptable photometry only down to $V\sim 18.5$. So the complete sample
(SGR12) is used only when the analysis is limited to the bright part of the
CMD, otherwise the best quality sample (SGR12R; 16992 stars) is adopted.   

\item SGR34, located at $(l;b)\sim(6.5;-16)$, sampling the second density clump
of the galaxy (see IGI-II). A total 
of 22603 stars have been measured, with a typical limiting magnitude of 
$V\sim 22.5$.

\item SGRWEST, located at $(l;b)\sim(5;-12)$, sampling a region near the edge
of the Sgr galaxy (as seen in the IWGIS isodensity map) toward the Galactic
Bulge. A total of 41462 stars have been measured, with a typical limiting 
magnitude of $V\sim 22.5$.   

\end{itemize}

The above fields are nearly oriented along the Sgr major axis (IWGIS), SGR34 is
$\sim 2 ~deg$ eastward from SGR12 and SGRWEST is $\sim 2 ~deg$ westward from 
SGR12.

A fourth control field of $24\times 9 ~arcmin^2$ (GAL) 
has been observed in a region
devoid of Sgr stars [$(l;b)\sim(354;-14)$]. The sample contains 8836
stars, with a limiting magnitude of $V\sim 22$.

The spatial distribution of stars in the SGDS fields is quite homogeneous and 
the crowding conditions are never critical. The completeness of the various 
samples is found to be very similar (see PAP-I).

\subsection{Reddening and degree of foreground contamination}

The interstellar extinction within each of the observed fields is rather 
homogeneous.
The reddening in the SGR34 field was assumed $E(V-I)=0.22\pm 0.04$,
as measured by MUSKKK (see PAP-I). 
Minor reddening differences were found between the SGR34 and the other SDGS
fields, and the corresponding corrections have been applied to report 
all the samples at the same reddening of the reference field SGR34. 
The extinction relations by Rieke \& Lebofski (1985; hereafter RL) has been 
always adopted both in PAP-I and in the present paper. 

\begin{figure*}
 \vspace{20pt}
\epsffile{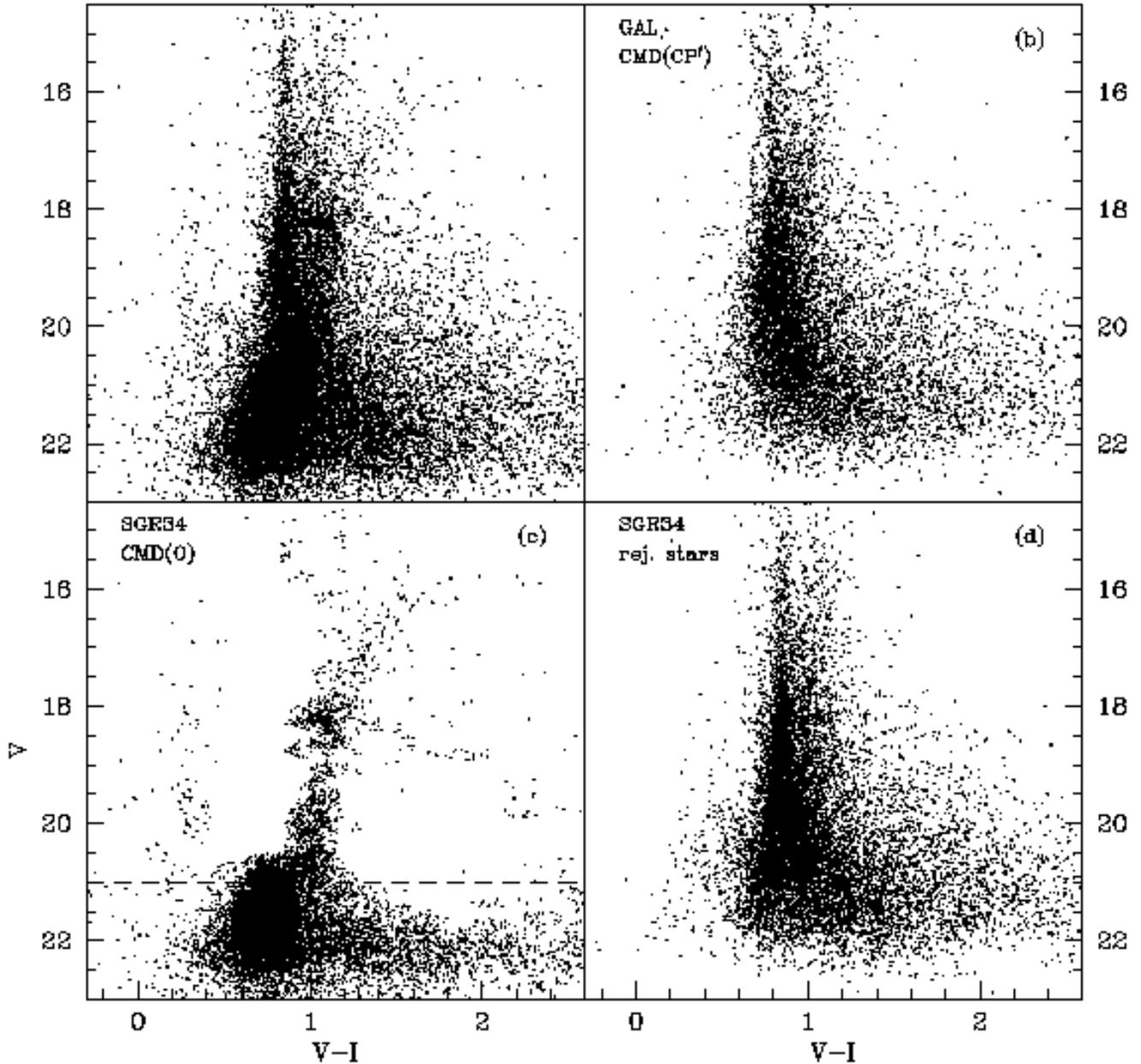}
 \caption{Decontamination of the SGR34 CMD(O+CP) [panel (a), 22559 stars] with 
 the GAL CMD(CP$\prime$) [panel (b), 8291 stars]. The decontaminated CMD(0) is
 shown in panel (c) (11036 stars) while the CMD of the stars rejected by the
 decontaminating algorithm, as putative CP members, is displayed in panel (d) 
 (11523 stars). 
  }
\end{figure*}

The various SDGS fields are affected by different amount of foreground star
contamination, since there is a strong density gradient in the Galactic Bulge
(and Disc, to a lesser extent) passing from $b=-16^o$ to $b=-12^o$. 
The degree of contamination was estimated with respect to the GAL field. 
It turned out that the density of stars belonging to the contaminating
population (CP) is nearly the same
for the SGR34 and the GAL fields, while is $\sim 1.6$ times higher in SGR12 and 
$\sim 3.6$ times higher in SGRWEST (see Sect. 3.2 for an account of the 
limitations of this approach).

\subsection{The two main stellar populations in Sgr}

The inspection of a typical Sgr CMD (MUSKKK, MAL, PAP-I) reveals that the
dominant stellar population is an intermediate-old one, with an extended RGB
and a clumpy red Horizontal Branch (HB). 
Different authors (MUSKKK, FAL, MAL) agrees that this
component is some Gyr younger than a typical globular cluster, the actual age
(and age difference) depending on the assumed metallicity (see PAP-I and Sect.
6). In PAP-I we referred to this main population as Sgr Pop A. 
It is still unclear
if Pop A is a {\em single age} population but it seems ascertained that a
metallicity spread is present (see PAP-I and references therein).
 
MUSKKK and MAL showed that a clear sequence of blue stars is also 
present above the MS Turn Off (TO) of Pop A, in the region of the CMD that is
often populated by Blue Stragglers (BS). Given the low stellar density in
the Sgr galaxy and the detection of Carbon Stars (IGI-II, WIC) 
this sequence (hereafter Blue Plume) has been interpreted
as an extended Main Sequence, signature of more recent star formation events.
In PAP-I we named this younger population Pop B and we showed that its spatial
distribution is similar to that of Pop A. 

\subsection{Analysis of non-decontaminated CMDs}

As we will show below, while statistical decontamination is in principle the
standard way to deal with strongly contaminated CMDs and has been successfully
performed in many cases (see Mighell et al. 1996 - hereafter MRSF - for example),
the actual application to the Sgr galaxy is very difficult. 
Up to now, the only attempt has been done by MUSKKK who provided
a {\em smoothed and integrated} (see MUSKKK) CMD of a ($15\times 15 arcmin^2$) 
field at the same position of SGR34.

On the other hand, all authors (MUSKKK, SL95, IWGIS, MAL) tried to take 
advantage of the fact that many important features of the Sgr CMD  
[as for example the brightest part of the RGB or 
the red HB] emerges sufficiently clearly from the sequences
produced by the contaminating foreground stars. 
Following this approach and comparing the Sgr CMDs to
the GAL one we obtained a number of results about distance, HB morphology etc. 
These results will be recalled below, when necessary.   

The most interesting result of PAP-I was the detection of a very metal poor 
component of the Sgr galaxy (similar to Ter 8, $[Fe/H] = -2.0$) 
that can be associated with a possible older population.

\section{Statistical decontamination of CMDs}

The basic idea of the Statistical Decontamination (SD) of a CMD is very
simple. We indicate with {\em O (Object)} the intrinsic population of the
studied stellar system and with CP the contaminating population. The observed
CMD [hereafter $CMD(O+CP)$] shows the contribution of both the intrinsic and
contaminating populations. The latter contribution to $CMD(O+CP)$ can be
estimated using a corresponding CMD of an adjacent field containing only 
foreground/background stars ($CP\prime$), under the assumption that it is
statistically representative of the contaminating population ($CP\prime \sim
CP$). The CP contribution on the $CMD(O+CP)$ is removed by comparing the local
density of stars in the two diagrams, $CMD(O+CP)$ versus $CMD(CP\prime)$: 

$$CMD(O+CP) - CMD(CP\prime) \sim CMD(O)$$

Obviously the technique can give optimal results when {\it (a)} the degree of
field contamination is moderate, i.e. the CMD(O+CP) is not dominated by
CP; {\it (b)} the CP$\prime$ sample is representative of the CP population
either in type and in density (see PAP-I), and {\it (c)} the Object features on
the CMD are intrinsically well defined and well populated, since a clear-cut 
local overdensity in CMD(O+CP) with respect to CMD(CP$\prime$) is the 
fundamental key for a star to enter in the CMD(O) sample.

As it shall be shown below, requirement (a) is only marginally fulfilled by 
the SGR34 and SGR12 samples while a good decontamination for SGRWEST results 
impossible due to overwhelming foreground contamination (see Sect. 3.2 and 4.3). 

In PAP-I we demonstrated that the GAL
sample is fairly representative of the CP present in the SGR34 CMD. Furthermore
we estimated the degree of contamination of the SGR12 and SGRWEST fields with
respect to SGR34. In the following we perform SDs scaling the CP$\prime$
local densities to the estimated degree of contamination.    

Concerning point (c), the low stellar density and the intrinsic metallicity
spread of Sgr conspire against the realization of a truly clean CMD(O), 
at least in the less populated sequences. However, thanks to the wide
SDGS samples, useful decontaminated CMDs have been obtained.

\subsection{The adopted algorithm}
  
In this paper we adopt an algorithm very similar to that described by MRSF.
Given a star belonging to $CMD(O+CP)$ and with coordinates [$V_{\star}\pm 
\sigma_v; (V-I)_{\star} \pm \sigma_{(V-I)}$] the number of stars falling into 
the ellipse of axis [$MAX(\sigma_V,0.1); MAX(\sigma_{(V-I)},0.1)$] around
($V_{\star}; V-I_{\star}$) on the original diagram ($N_{O+CP}$) and on the 
``pure foreground'' diagram ($N_{CP\prime}$) are computed. 
The actual dimension of the cell is an important parameter since
if the cell is too small then only few stars fall into and the derived 
probability would be of little statistical significance. This occurrence is
prevented by setting a fixed minimum dimension of the cell. 

On the other hand, if
the cell is too large the local number star density can be influenced by stars
belonging to a feature physically distinct to the one to which the considered
star belongs. The effect can become very destructive in the faint part of a CMD
where $\sigma_V$ and $\sigma_{(V-I)}$ are larger. To minimize the problem we
weighted each star present in the cell with an elliptical bi-dimensional Gauss 
distribution, according to its distance from the center of the cell, so that a
star in the vicinity of ($V_{\star}; V-I_{\star}$) gives a greater contribution
to $N_{O+CP}$ or $N_{CP\prime}$ with respect to a star which fall near the edge
of the cell itself. Many test (either on SDGS CMDs or on CMD of globular
clusters artificially contaminated) showed that much cleaner and {\em stable}
CMD(O) are obtained with this kind of cell than with a crude rectangular
one. While in the case of very well defined Object loci and low contamination
the two approaches can give equivalent results, in more difficult cases (as the
present ones) the ``weighted cell'' technique is clearly preferable.   
 
According to MRSF, the probability $P$ that a given star in the CMD(O+CP) is 
actually an Object member is defined as follow:

\begin{equation}
P = {1 - MIN({{\alpha ~N_{CP\prime}\over{N_{O+CP}}}, 1.0})}
\label{eq1}
\end{equation}

where $\alpha$ is the ratio of the area of the $O+CP$ field to the area of the 
$CP\prime$ field. Once $P$ is calculated for a given star, it is compared with
a randomly drawn number $0<P\prime<1$ and if $P\prime\le P$ the star is
accepted as an Object member, otherwise it is rejected and considered as a
CP member (see MRSF for a more detailed description).

\begin{figure*}
 \vspace{20pt}
\epsffile{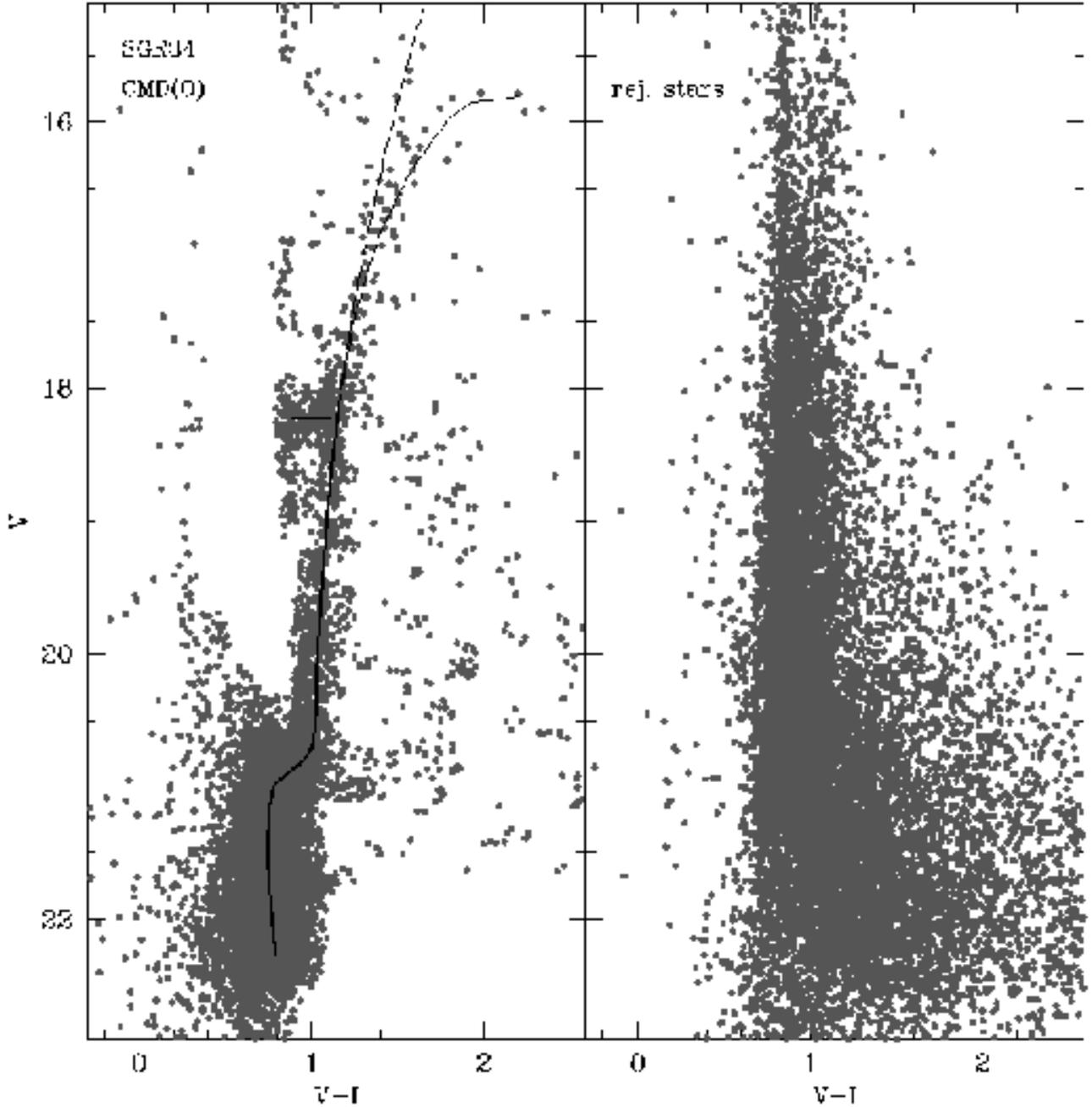}
 \caption{Decontamination of the SGR34 CMD(O+CP) [22559 stars] with 
 the MCF CMD(CP$\prime$) [13165 stars]. The decontaminated CMD(0) is
 shown in the left panel (9187 stars) while the CMD of the stars rejected by 
 the decontaminating algorithm, as putative CP members, is displayed in the
 right panel (13372 stars). The faint parts of the decontaminated CMD presented
 here are clearly cleaner with respect to the CMD shown in the panel (c) of Fig.
 1. The Pop A ridge line (see tab. 1) is superimposed to the CMD. Two different
 ridge lines are adopted for $V<18$ to account for the observed spread of the
 RGB stars.}
\end{figure*}

\subsection{Applications}

In the following decontaminations of the CMDs the parameter 
$\alpha$ has been set to 
the appropriate area ratio. Furthermore we used this parameter to account for
the different density of the CP population in the different CMDs (SGR34, SGR12
and SGRWEST). This ``CP density tuning'' was obtained by increasing
the $\alpha$ ratio by a factor which accounts for the different CP density with
respect to our $CP\prime$ sample GAL, according to the estimates of PAP-I.
Being $\alpha$ the pure area ratio, let's call $\alpha^{'}$ the final parameter
adopted. For each considered field we determined $\alpha^{'}$ as follows:

\begin{itemize}

\item For SGR34: $\alpha^{'}=1.0 \times \alpha$

\item For SGR12: $\alpha^{'}=1.6 \times \alpha$

\item For SGRWEST: $\alpha^{'}=3.6 \times \alpha$

\end{itemize} 

It is important to stress that this approach is based on the implicit assumption
that the CP in the various SDGS fields differs only in density, which in 
general is not true. 
The ``CP density tuning'' factors were derived from the normalization
of star counts in a narrow box of the CMD in which the foreground population is 
dominant (see PAP-I). Changes in the stellar mix (for example in the ratio 
between bulge and disk stars) cannot be accounted for, and may lead to the 
adoption of density tuning factors which are not adequate for other regions of 
the CMD. Some of the results presented in this paper may be affected by this
additional uncertainty (in particular those presented in Sect. 5.1). 

The rationale of the adopted decontamination strategy has been widely discussed
in PAP-I, together with {\em pros} and {\em cons} of possible alternative
approaches. Suffice to say here that the present application is optimized to
recover the Sgr features on the CMD that are more vexed by foreground
contamination, i.e. the upper MS, the SGB and the lower RGB (see Fig. 1).     

In Fig. 1 the result of the decontamination of the SGR34 CMD using GAL as 
$CP\prime$ is shown. Panel (a) reports the original CMD(O+CP), panel (b) the
CMD($CP\prime$) - i.e. the GAL CMD -, in panel (c) the decontaminated CMD(O)
is presented and, finally, the CMD of the stars rejected by the subtraction
algorithm (as putative CP members) is reported in panel (d). The horizontal
line in panel (c) shows the level above which a reliable correction for
completeness can be performed.

\begin{table}
 \centering
 \begin{minipage}{140mm}
  \caption{SGR34 Pop A - Ridge line.}
  \begin{tabular}{@{}lccclc@{}}
    V    &     V-I      &      V-I      &       V-I     & V & V-I\\
         & for $V\ge18$ & RGB {\em red} & RGB {\em blue}& HB& HB \\
         &              &               & 	        &   &	   \\
 15.14   &  -           & -             & 1.645 	&- &-	   \\
 15.5    &  -           & -             & 1.569 	&- &-	   \\
 15.80   &  -           & 2.202         & -             &- &- \\
 15.82   &  -           & 2.104         & -             &- &- \\
 15.89   &  -           & 1.880         & -             &- &- \\
    16   &  -           & 1.783         & 1.463         &- &- \\
  16.5   &  -           & 1.521         & 1.376         &- &- \\
    17   &  -           & 1.354         & 1.298         &- &- \\
  17.5   &  -           & 1.247         &  1.23         &- &- \\
    18   & 1.176        & 1.176         & 1.176         & 18.23 & 0.876\\
  18.5   & 1.126        & -		& -		& 18.23 & 1.116\\
   19	 & 1.088        & -		& -		&- &- \\
  19.5   & 1.06         & -		& -		&- &- \\
 20.05   & 1.034        & -		& -		&- &- \\
 20.55   & 1.017        & -		& -		&- &- \\
 20.75   & 0.980        & -		& -		&- &- \\
  20.8   & 0.955        & -		& -		&- &- \\
 20.85   & 0.908        & -		& -		&- &- \\
  20.9   & 0.860        & -		& -		&- &- \\
 20.95   & 0.810        & -		& -		&- &-\\
    21   & 0.780        & -		& -		&- &-\\
 21.16   & 0.753        & -		& -		&- &-\\
 21.26   & 0.751        & -		& -		&- &- \\
 21.43   & 0.749        & -		& -		&- &- \\
 21.50   & 0.744        & -		& -		&- &-\\
 21.57   & 0.751        & -		& -		&- &- \\
 21.67   & 0.756        & -		& -		&- &- \\
 21.89   & 0.764        & -		& -		&- &-\\
 22.28   & 0.794        & -	        & -	       & - &-\\
\end{tabular}
\end{minipage}
\end{table}

It results clear from Fig. 1 that the SD was largely successful.
The {\em globular cluster -like} shape of Sgr Pop A clearly emerges in panel 
(c), the upper MS and the Sub Giant Branch (SGB) are now
clear, while they were mostly obscured by CP stars in the original
CMD. On the contrary, much of the residual contamination below $V\sim 20.5$ is
probably due to the relatively low limiting magnitude of the GAL sample
(see PAP-I).

The effect of SD on the local density of stars in the various
features of the final CMDs is hard to evaluate, also in the region of the CMD
where completeness corrections can be performed. So, a detailed
Luminosity Function could be affected by artificial bumps and gaps on small
scales. Star counts remain a useful investigation tool when specific test are
applied (as those of PAP-I), i.e. counting stars in relatively large boxes
around well populated features. 
On the other hand the SDs make available in a much clearer form the
informations contained in the {\em shape} and {\em position} of the various
CMDs features, i.e. - mainly - informations about age and metallicity.

In PAP-I it was suggested that the deeper CMD of the MUSKKK Control Field
(hereafter MCF) can be used for decontamination of the SDGS diagrams, since 
the position of the field is the same of GAL. Few experiments showed that a
superior decontamination of the faint part of CMDs can be obtained using MCF
instead of GAL. In the following analysis MCF is adopted as the standard
decontamination sample. The CMD shown in Fig. 1  (and the other obtained for
SGR12 and SGRWEST using the GAL control field\footnote{Not shown here for
obvious reasons of space}) will be used as
independent check of possible spurious features in the obtained CMD(O) 
due to small inhomogeneities between the SDGS and the MUSKKK control field 
samples.
In Fig. 2 the results of the decontamination of the SGR34 CMD with the MCF CMD
are presented [left panel: CMD(O); right panel: CMD of the rejected stars].

\begin{figure*}
 \vspace{20pt}
\epsffile{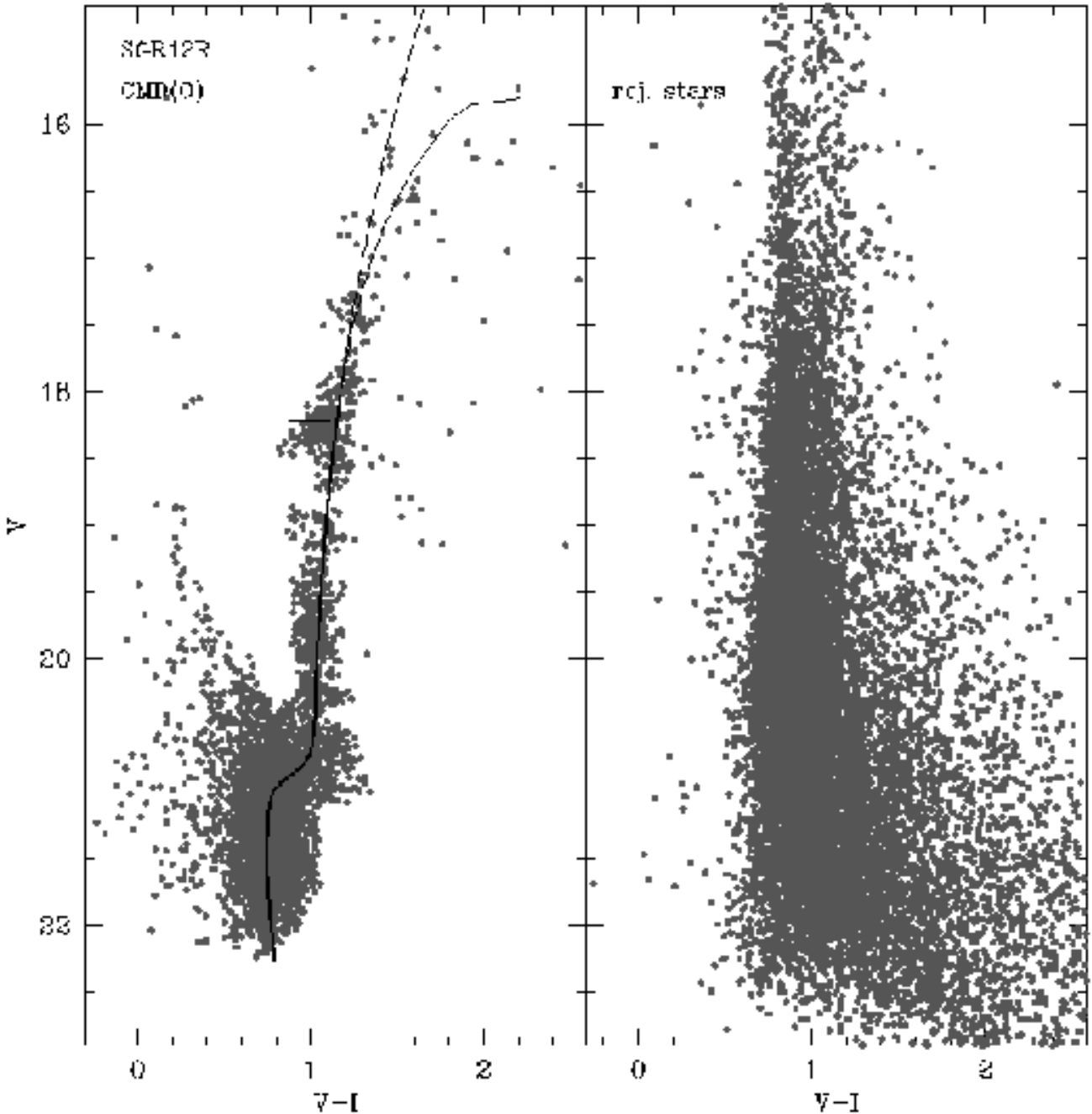}
 \caption{The same as Fig. 2 for the SGR12R sample (left panel: 4017 stars,
 right panel: 12936; total number of stars in the original CMD: 16953).  
 The SGR34 ridge line has
 been superposed to allow a comparison between the main features of the CMDs
 in the two fields. The ridge line provides an excellent fit to the data 
 (particularly in the TO region).}
\end{figure*}
 
There are features common to all the decontaminated CMDs presented in this
paper that deserve some comments.
First, all the stars fainter than $V\sim
17$ and redder than $(V-I)\sim 1.2$ are obviously not members of Sgr, being
mere residuals of the decontamination. 
Second, any apparent structure along the RGB (like gaps or small horizontal
clumps) has to be considered as an artifact of the decontamination process. 

\section{Statistically decontaminated CMDs: comparisons between SDGS fields}

Since SGR34 is the SDGS field which couples good photometric quality and the
lowest degree of foreground contamination we first discuss the
CMD(O) obtained from this field and then we will analize the SGR12 and SGRWEST
samples by comparison with SGR34. We will concentrate on the ``globular cluster
like'' bulk of the Sgr population (Pop A). Pop B will be discussed in 
Sect. 6.4. 

\subsection{The SGR34 field}

The decontaminated CMD of the SGR34 field is very similar, in shape, 
to that of a typical (metal rich) globular cluster, with a faint TO region, 
an extended RGB and a stubby red HB around $V\sim 18$. 

The average ridge line of the decontaminated SGR34 CMD (Pop A) is 
also presented in the left panel of Fig. 3 and in table 1. The line has
been obtained by calculating the mean color in bins encompassing 0.1 mag from
$V=22$ to $V=20.5$ and 0.5 mag from $V=20.5$ to $V=18$, and then applying a
{\em 2-$\sigma$ clipping} algorithm similar to that described by Sarajedini \&
Norris (1994). The spread of the stars in the brightest part of the RBG
is such that an {\em
average} ridge line would have little sense. We provide two different ridge
lines for $V<18$, representing the conservative range of RGB stars distribution
in this portion of the CMD. In table 1 the two ridge lines adopted for $V<18$
are reported as RGB {\em blue} and RGB {\em red} according to their relative
position in the CMD. Despite the large spread of the RGB, the 
Upper MS + SGB sequences are relatively narrow, very similar to
those of a genuine globular cluster observed with the same photometric
precision (see the CMD of Ter 8 by MoAL, for instance). The ``thin'' SGB is
particularly outstanding considering the photometric errors involved and the
effects of the decontamination process.

Many ``artificial decontamination'' tests convinced us 
that the net effect of the decontamination process on a typical SGB is to widen
it and that only {\em very sparse} sequences can be accidentally removed 
from the CMD by the cleaning process.  
So, we are forced to conclude that the large
majority of Pop A stars {\em share the same SGB sequence} within the
photometric errors. This statement can have far reaching consequences, as we
discuss in Sect. 6.

Sparse groups of stars appears nearly vertically aligned around 
$(V-I)\sim 0.9$, from $V\sim 19$ to $V\sim 15$. 
This feature is not present in the analogous CMD(0) obtained using
GAL as $CP\prime$ (see Fig. 1c) and consequently have to be
considered as a spurious residual of the actual decontamination process.

On the other hand the presence of the above quoted Blue Plume and its
association with the \sgr has been robustly assessed in PAP-I, so it is no
surprise to find it clearly emerging around $[V\sim 20.5; (V-I)\sim 0.4]$.
Also this feature has an obvious counterpart in many galactic globulars as the
well know BS sequence 
(see Ferraro, Fusi Pecci \& Bellazzini 1995
and Bailyn 1995, for references), which is observed also in clusters of very
low stellar density (e.g. NGC 5053, Nemec \& Cohen 1989). The main reason to
associate the Sgr Blue Plume with the upper Main Sequence of a younger
population (Pop B) is the detection of Carbon Stars (IWGIS, Whitelock, Irwin \&
Catchpole 1996; see also PAP-I) which unequivocally points to the presence of 
younger stars in the galaxy (e.g. Da Costa 1998). 
However it cannot be
firmly established, at present, if all the Blue Plume stars are progenitors of
the observed Carbon Stars or if a (perhaps significant) fraction of them is
constituted by genuine BS stars. 

If the Blue Plume is indeed an extended MS one would expect to observe
also some SGB star associated with it, between the red edge of the Blue Plume
itself and the base of the RGB of Pop A, while none is observed. Here we are
facing the opposite situation with respect to the Pop A SGB (see above): Pop B
SGB stars would be so few and so sparsely distributed in the CMD that would
provide no significant local overdensity over the CP in this region, and
probably have all been ``erased'' by the decontamination process. This fact
would strongly lower our capability to constrain the SFH of Pop B (see Sect.
6.2).

\subsection{The SGR12 field}

The decontaminated CMD of the SGR12R sample is shown in Fig.3 (left panel)
together with the CMD of the rejected stars (right panel). The SGR34 ridge 
line already presented in Fig. 2 is superimposed to the decontaminated diagram 
to allow a direct comparison between tha main Pop A features in the two fields.

The similarity is indeed remarkable, and the SGR34 ridge line fits very well
all the observed SGR12R features.  
The only noticeable difference is the presence of stars significantly
redder than the SGR34 ridge lines near the RGB tip, and the apparent 
redder HB morphology in the SGR12R sample with respect to the SGR34 one. 
Both differences have been already noted in PAP-I and tentatively interpreted 
as indications of the presence of more metal rich stars in the region of the 
Sgr galaxy covered by the SGR12 field (i.e., the center of density; see also 
SL95 and MAL). 
The difference in the
distribution of RGB stars will be analyzed in detail in Sect. 5. Here we
briefly discuss the difference in HB morphology.

Since the HB of the decontaminated SGR34 sample shown in Fig. 2 merges at the
blue end with the spurious vertical sequence described in the previous section,
we compare the HB morphology of the SGR34 and the SGR12R samples decontaminated 
with our own Control Field (GAL), including completeness corrections (see
PAP-I). 

\begin{figure}
 \vspace{20pt}
\epsffile{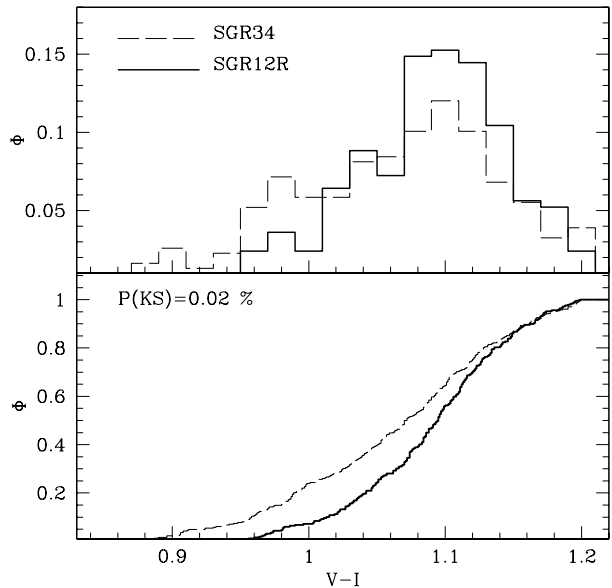}
 \caption{Upper panel: $V-I$ color histograms (normalized to the total number of 
 stars) of the HB stars for the SGR12R (bold continuous line) and the SGR34 
 (thin dashed line) samples. Lower panel: the cumulative distribution for the
 same stars. The probability that the two samples are extracted from the same
 parent population (according to a Kolmogorov-Smirnov test) is also reported on
 the upper left corner of the diagram.}
\end{figure}

In the upper panel of Fig. 4 the $V-I$ histogram 
(normalized to the total number of stars) of the stars having $17.5<V<18.5$ and
$0.55<V-I<1.2$ (see PAP-I) are reported for the SGR34 sample (thin dashed line) 
and for the SGR12R sample (bold continuous line). 
While the two distributions peak at the
same color bin, the fraction of stars redder than $V-I=1.0$ is significantly
higher for SGR12R, as can be readily appreciated from the cumulative
distribution shown in the lower panel of Fig. 4. A Kolmogorov-Smirnov test (a
test particularly well suited to analyze {\em shifts} between distributions)
shows that the probability that the two samples are extracted from the same
parent population is only $0.02 \%$.

The analysis of the decontaminated CMD confirms the results of PAP-I: the
HB morphology of the main population in the SGR12 field is slightly redder
than in SGR34. It is important to stress that the present conclusion
concerns the red part of the Sgr HB. To have a complete view of
the morphology one have to take into account also RR Lyrae stars, which are
certainly present (Mateo et al. 1995b, Alard 1996, Alcock et al. 1997), and
blue HB stars, whose possible presence have been suggested in PAP-I.   
 
\subsection{The SGRWEST field}

As remarked in Sect. 3, the success of a statistical decontamination strongly
depends on the relative surface density of CP and Object stars present in the
observed field. The usual applications of the SD technique consists in the 
removal of a minority of field stars from the CMD of a populous cluster (see
, for example, MRSF and Olsen et al. 1998).

From the performed decontaminations it results that even in the
less contaminated SDGS fields the majority of stars belong to the
foreground population: $41 \%$ of stars are accepted as Sgr members in the
SGR34 field and only $24 \%$ in the SGR12R one. 

Since, as shown in PAP-I, the surface density of CP stars grows significantly
toward low galactic latitudes, it is not surprising that a satisfying 
decontamination cannot be obtained for the SGRWEST CMD. The solution obtained
following the prescriptions of Sect. 3 fails to remove all of the CP main 
sequence from the CMD, while it removes part of the Sgr RGB instead. The final
decontaminated CMD can't allow further analysis and simply confirm the results
obtained in PAP-I from the non-decontaminated CMD, i.e. {\em (a)} Sgr stars 
are present in the SGRWEST field, and {\em (b)} the overall CMD morphology is
very similar to that observed in the other SDGS fields.

\section{Metallicity}

In PAP-I we presented a preliminary comparison of the observed RGBs with
the ridge lines of template galactic globulars from DCA90. Now we can
perform a more detailed analysis on the decontaminated CMDs, using ridge lines 
derived from more recent
photometry and covering the CMDs of the template clusters from the RGB tip to - at
least - one magnitude below the MSTO. This will allow us to derive 
metallicity and age estimates with {\em the same} templates, with a fully
consistent procedure.

\begin{figure*}
\vspace{20pt}
\epsffile{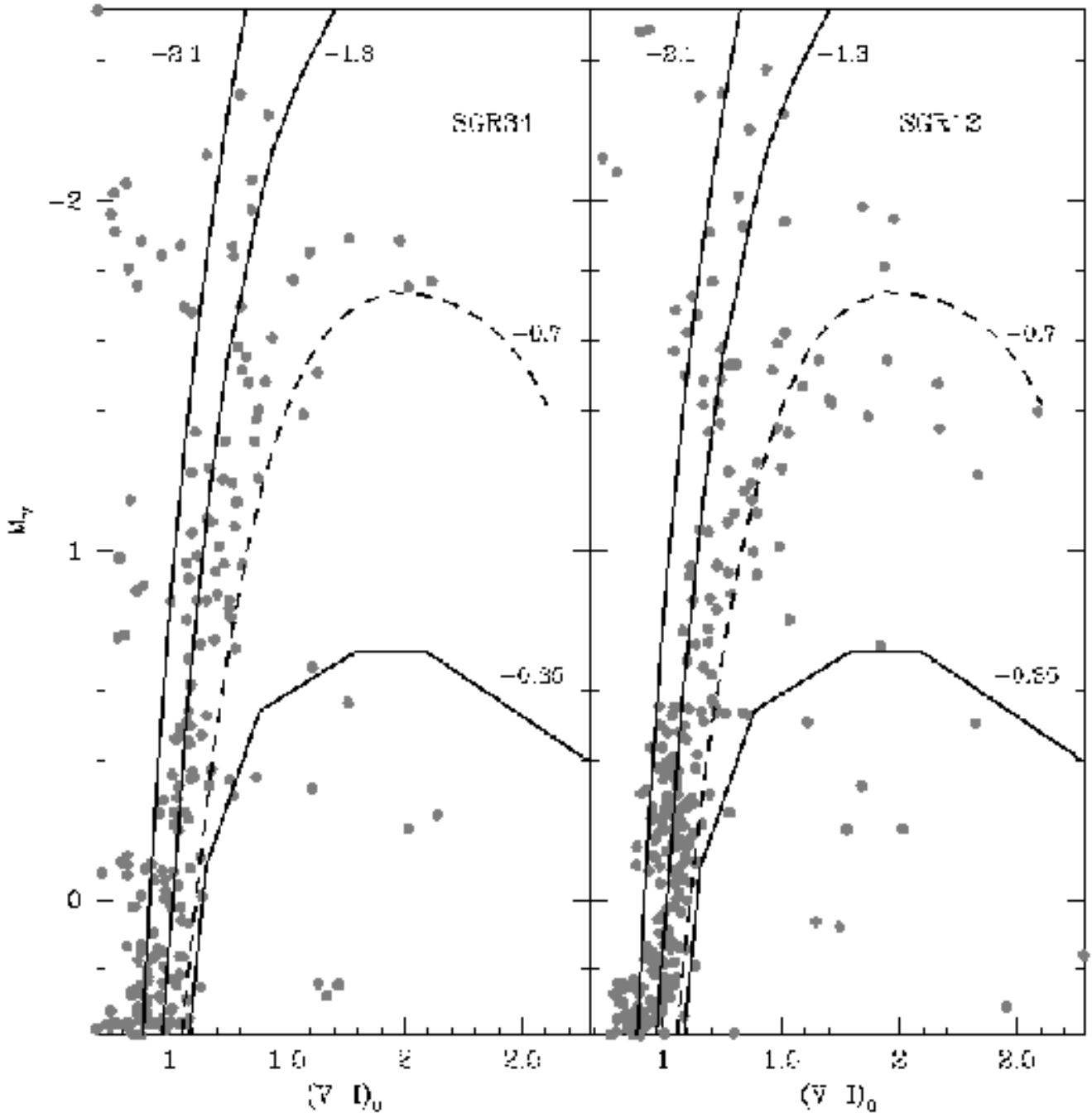}
\caption{The bright part of the decontaminated CMD of SGR34 (left panel), and
SGR12 (right panel) reported in the
absolute plane. The superposed lines are the ridge lines of  
M68 ($[Fe/H=-2.1$), M3 ($[Fe/H=-1.34$), 47 Tuc ($[Fe/H]=-0.71$), and
NGC6528 ($[Fe/H=-0.35$), from left to right. 
The labels near the tip of the lines are the quoted
[Fe/H] values.}
\end{figure*}

The adopted ridge lines are taken from the following templates:

\begin{enumerate}

\item M68 ($[Fe/H=-2.1]$) from Walker (1994). The distance modulus was derived
adopting his value of $<V(RR Lyrae)>$ and applying the 
$M_V(RR Lyrae) ~vs. ~[Fe/H]$ relation by DCA90. From his estimate of the
interstellar extinction $E(V-I)=0.09$ has been obtained and finally 
$\mu_0=14.95$ has been adopted.

\item M3=NGC6752 ($[Fe/H=-1.34$)
from Ferraro et al. 1997 adopting their estimates $\mu_0=15.05$ and 
$E(V-I)=0.0$. 
The ridge line of the MS+SGB in the (V;V-I) plane was 
not published by Ferraro et al. 1997 and has been derived directly from their 
data and joined with their RGB ridge line.

\item 47 Tuc=NGC104 ($[Fe/H]=-0.71$) derived from the photometry of
Kaluzny et al. (1998) adopting $\mu_0=13.51$ and $E(V-I)=0.064$ according to
DCA90. The photometry of Kaluzny does not cover the brightest part of the RGB.
However the ridge line of DCA90 smoothly joints the one described above at
$M_V\sim 1.1$ and provides the ideal complement for the bright RGB. 

\item NGC6528 ($[Fe/H]=-0.35$) derived from the data of Ortolani et al.
(1995). Adopting $\mu_0=14.45$ and $E(V-I)=0.68$, according to Bruzual et al.
(1997), a significant mismatch is noted between the position of the HBs of SGR34 
and NGC6528. A fair match is obtained if $E(V-I)=0.83$ is assumed, in
good agreement with the alternative value proposed as by Bruzual et al. (1995),
$E(V-I)=0.81$ (see their Note 2). 

A further template will be used in Sect. 6, for the age analysis:

\item Pal 1 ($[Fe/H]=-0.7$) from Rosenberg et al. (1998a hereafter R98; see 
also Rosenberg et al. 1998b for the spectroscopical metallicity measurement), 
adopting their $E(V-I)=0.20$. R98 derived $\mu_0=15.25\pm 0.25$ 
by comparison with the ridge line of 47 Tuc by Hesser et al. (1987) and 
adopting $\mu_0=13.35$ for this latter cluster. Since here we have always
adopted $\mu_0=13.51$ for 47 Tuc (also in deriving the distance modulus of Sgr,
see PAP-I) we properly scaled the R98 distance modulus, obtaining
$\mu_0=15.41\pm 0.25$ for Pal 1. In this way we set a fully (internally)
consistent distance scale for Sgr, Pal 1 and 47 Tuc. The adoption of any
alternative distance modulus for 47 Tuc would imply a corresponding 
correction in the moduli of Pal 1 and Sgr and the relative position of the ridge
lines on the CMD would be unaffected.    

The ridge line of Pal 1 is defined only up to the HB level. 
However the
cluster has the same metallicity of 47 Tuc and the DCA90 ridge line for this
cluster (represented in the plots as a dashed line) matches the line 
by R98 (continuous line) around $M_V=0.75$, with the 
quoted distance and reddening assumptions, so it is adopted as the bright 
complement to the Pal 1 ridge line.  

\end{enumerate}
 
In Fig. 5, the bright part of the decontaminated CMD of SGR34 (left panel) 
and SGR12 (right panel) are 
reported in the absolute plane ($M_V; (V-I)_0$) by adopting $\mu_0=17.25$, 
$E(V-I)=0.22$\footnote{It has to be recalled that differential reddening 
(and extinction) correction has already been applied to the SGR12 and SGRWEST
samples to report them at the same reddening than SGR34 (see Sect. 2.2 and
PAP-I). So the total correction to report the former two samples 
into the ($M_V$,$(V-I)_0$) plane are $E(V-I)=0.23$ for SGR12, and 
$E(V-I)=0.25$ for SGRWEST.}. The ridge lines of M68, M3,
47 Tuc and NGC 6528 are superimposed on the CMDs and the corresponding [Fe/H]
values are marked near the ridge lines tips.
The left panel of Fig. 5 shows that the bulk of the SGR34 RGB stars
is nicely bracketed between the ridge lines of M3
and 47 Tuc. The distribution appears skewed toward the M3 ridge line and Sgr
``putative member'' stars bluer than this line are also present. 
No evident discontinuity in the distribution of stars between the 
quoted ridge lines is evident. 

Interpreting the spread
of the upper RGB as a spread in metal content between Sgr stars, 
we have that the majority of Sgr stars have $-1.3\le [Fe/H] \le -0.7$,
but the metallicity distribution probably reach a metallicity as low as
$[Fe/H]=-2.1$ (as shown also in PAP-I). Note that the global spread induced by
photometric errors at the considered magnitude level is negligible (see PAP-I),
so we are observing a physical spread in color.

The region of the CMD immediately below the almost vertical extension of the
47 Tuc ridge line ($1.5<(V-I)_0<2.3$ and $M_V>-1.6$) is remarkably devoid of
stars. This fact suggests $[Fe/H]=-0.7$ as a firm upper limit to Sgr metallicity
in this region of the galaxy. At $M_V\sim -0.7$ and
$1.2<(V-I)_0<1.8$ there are a few
stars clustered around the ridge line of NGC6528 that could be tentatively
associated with a sparse and very metal rich population. 
Most probably, they are
spurious residuals of the decontamination as the stars with the
same colors at $M_V>0$. 

The SGR12 case (Fig. 5, right panel) is very similar to the SGR34 one. 
The only apparent difference stands in the significant number of stars
redder than the 47 Tuc ridge line, around $M_V=-1.5$. The metallicity spread of
stars in this field seems larger, with respect to SGR34, and the upper limit
metallicity somewhat higher than $[Fe/H]=-0.7$, but still $[Fe/H]<-0.35$.

Evidences for a significant metallicity spread in Sgr Pop A has been already 
pointed out by many authors (MUSKKK, SL95, MAL) and appears to be confirmed by
very recent high resolution spectroscopy of few Sgr giants by Smecker-Hane et
al. (1998), who find $-1.5\le [Fe/H] \le +0.11$ (see Sect. 4.3 and also 
Pasquini et al. 1999).

\subsection{Metallicity distribution} 

Since a significant metallicity spread is evident in all the observed fields, it
would be of paramount importance to know the {\bf metallicity distribution} of
Pop A stars. It is obvious that the right procedure to obtain such a 
distribution would imply accurate spectroscopic analysis that would lead
simultaneously to a selection of {\em true member stars} and to a metallicity
estimate {\em of each single measured star}. However this kind of project
implies a huge observational effort, since to collect a statistically
significant sample ($\sim 50$) of Sgr members in just one of the SDGD 
fields, some 80-100 individual spectra have to be obtained (see Sect. 4.3). 
On the other hand the SDGS provides a total of more than 150 upper RGB 
stars ($\sim 50$ per field) suitable for some photometric metallicity estimate. 
Their membership cannot be stated on a star by star basis but the 
{\em statistical membership} of the sample is robustly stated, since the upper
part of the RGB is not seriously contaminated.

We obtained a coarse metallicity distribution of stars in each of the SDGS
fields by the following steps:

\begin{itemize}

\item Only the portion of the CMDs of Fig. 5 delimited by  
$(V-I)_0>0.9$ and $-2.5<M_V<-0.8$ has been considered, in order to
avoid contamination of the sample by red HB stars or spurious residuals by the
decontamination processes at very blue colors or too faint magnitudes.

\item We counted the number of stars (1) bluer than the M68 ridge line,
(2) lying between the M68 and the M3 ridge lines, (3) lying between the M3 
and the 47 Tuc ridge lines, and (4) redder than the 47 Tuc ridge line. 
The described bins are illustrated in Fig. 6.

\begin{figure}
\vspace{20pt}
\epsffile{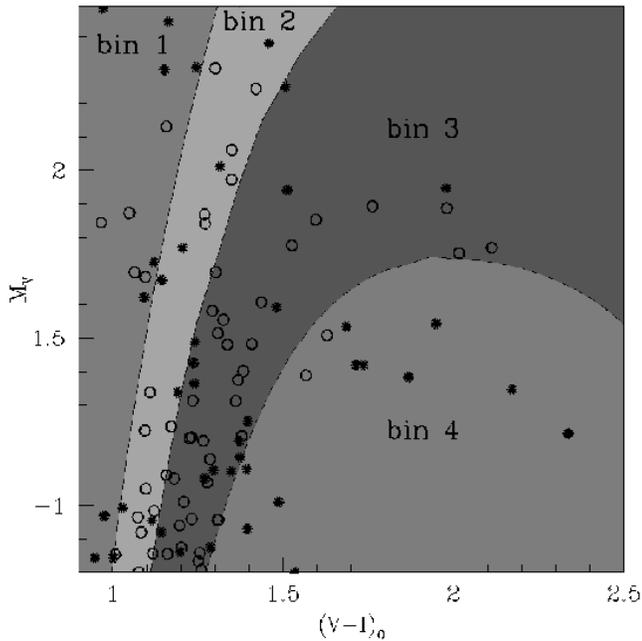}
\caption{Metallicity bins used to derive the metallicity distribution
displayed in Fig. 7 . The ridge lines are the same of Fig. 5. 
Bin 1 corresponds to $[Fe/H]<-2.1$, bin 2 to $-2.1\le [Fe/H]<
-1.3$, bin 3 to $-1.3\le [Fe/H]< -0.7$, and $[Fe/H]\ge -0.7$.
Open circles are from the SGR34 sample, asterisks from the SGR12 sample.}
\end{figure}

\item The above defined bins were interpreted as coarse metallicity bins,
according to the metallicity of the templates and in Fig. 7 we report the
resulting histograms (upper panel: SGR34, lower panel: SGR12). 
The bin 1 and 4 have been devised to detect the limits of
the metallicity distribution: a significant decrease of frequency in this bins
strongly indicates respectively $[Fe/H]=-2.1$ and $[Fe/H]=-0.7$ as the true 
limits of the distribution. The left side of bin 1 and the right side of bin 4
histograms have been represented with dotted lines to indicate that, if a
significant number of stars is indeed present in these bins a well defined 
metallicity limit cannot be assumed.

\end{itemize}

\begin{figure}
\vspace{20pt}
\epsffile{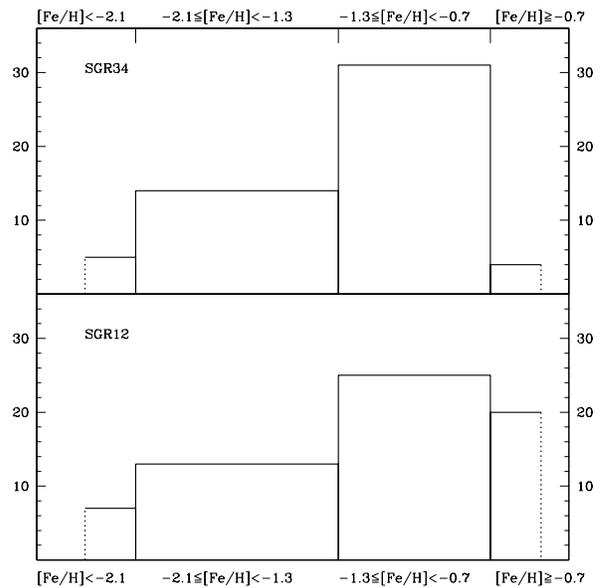}
\caption{Coarse metallicity distributions in the SDGS fields from the
distribution of stars in the upper RGB region of the CMDs.}
\end{figure}

The two histograms are rather similar. The lower metallicity limit appear to be
[Fe/H]=-2.1 in both samples, confirming the results of PAP-I. 
The peak of the distribution is clearly in bin 3. 
The number of stars between $-1.3\le [Fe/H] < -0.7$ is 
double than that in the $-2.1\le [Fe/H] < -1.3$ bin. According to Fig. 7 it
can be concluded that the Sgr galaxy produced more than $65 \%$ of Pop A stars
at metallicity $[Fe/H] \ge -1.3$, the very metal poor stars (around
$[Fe/H]=-2.$) being a minor component (see Pap-I).

The SGR34 distribution has a defined strong metallicity limit at $[Fe/H]=-0.7$
while the $30 \%$ of SGR12 stars have $[Fe/H]\ge-0.7$.   

Fig. 7 shows for the first time that {\em the Sgr Pop A
metallicity distribution presents significant structures}, certain metallicity
ranges being more populated than others. In Sect. 6 we will show that this
{\em can be interpreted as a variation of the Star Formation Rate (SFR) with
time}.  

\section{The Age of Pop A}

The first age estimate for the Sgr Pop A was made by
MUSKKK, who fitted their decontaminated CMD of the SGR34 region with an 
isochrone of metallicity $[Fe/H]=-1.2$ and age 10 Gyr. 
They concluded that the bulk of the Sgr
stars are significantly younger than typical Galactic globulars but older than
intermediate age dSphs, as Carina. Subsequently, Mateo et al. (1996) studied a
sample of Sgr stars projected in the same direction of the M55 globular
cluster. They fitted the observed CMD (which is severely contaminated by M55
stars) with an isochrone of $[Fe/H]=-0.5$ and age 12 Gyr. 

Fahlman et al. (1996) also found evidences for the presence of Sgr stars in a
field toward M55. The sparse Sgr MS in their CMD was equally well
fitted by two different isochrones: $[Fe/H]=-0.79; ~age=10 Gyr$ or
$[Fe/H]=-1.27; ~age=14 Gyr$. Using the same set of isochrones they found an age
of 16 Gyr for the Galactic globular clusters M55 and M4 and so confirmed a
younger age for Sgr with respect to classical globulars.

While all the above determination of the age of Pop A rely on isochrone fitting
(a technique often leading to ambiguous results and far from optimal in
deriving age differences), MAL showed that the main CMD loci of Sgr are nearly
coincident with those of the globular cluster Ter 7 ($[Fe/H]\sim -1$ and $\sim
4-6 ~Gyr$ younger than typical Galactic globulars of similar
metallicity\footnote{The actual value of the metallicity of Ter 7 is subject of
debate; see Fusi Pecci et al. 1995, Da Costa \& Armandroff 1995 and Sarajedini 
\& Layden 1997, for discussion.}). 
This result provided (1) an empirical evidence of
the younger age of Pop A with respect to standard old Galactic globulars and
(2) the proof that the formation of globular clusters in Sgr ended with the
star formation event that originated Pop A (see MoAL).

\subsection{Distance and reddening -independent age estimates}   
    
MoAL presented the Age-Metallicity Relation (AMR) of the globular cluster 
system of
the \sgr, based on the measured $\Delta V^{TO}_{HB}$ 
(the difference between the HB level and the magnitude of the TO point) and an 
$age=f(\Delta V^{TO}_{HB},[Fe/H])$ equation taken from Chaboyer, Demarque \&
Sarajedini (1996). They showed that, (1) the oldest 
clusters of the Sgr system (Ter 8 and M54) are coeval with the oldest 
Galactic globulars, (2) there is a significant range in age between the metal 
poor globulars of the \sgr (i.e. Arp 2 is nearly 4 Gyr younger than Ter8 and 
M54) and (3) a significant chemical enrichment occurred in the time between the
formation of Arp 2 ($[Fe/H]=-1.84$) and the formation of Ter 7 ($[Fe/H]\ge
-1.$), some 2-3 Gyr after.

For SGR34 we measured $V(TO)=21.5\pm 0.1$ and $V(HB)=18.23\pm 0.02$ (PAP-I),
$\Delta V^{TO}_{HB}=3.27 \pm 0.1$ is derived, which places Sgr Pop A between
Arp 2 and Ter 7 in the AMR of MoAL (see their Fig. 8), depending on the assumed
metallicity. It is worth noting that the adopted $V(HB)$ is the mean V
magnitude of the red HB of Sgr, while the correct HB level should be  
the mean magnitude of the HB in the RR Lyrae
region of the branch. This is expected to be fainter than $V(RedHB)$ by 
$\sim 0.05 - 0.15$ (see PAP-I), so the quoted $\Delta V^{TO}_{HB}$ 
value can be regarded as an upper limit. Taking this correction into
account the $\Delta V^{TO}_{HB}$ of SGR34 is found to be virtually equal to
that of Ter 7, confirming the result of MAL.

The observables to estimate the relative ages of globular clusters are
typically classified as {\em vertical} - when based on the {\em luminosity} of 
the TO points (as $\Delta V^{TO}_{HB}$) - or {\em horizontal}, when based on 
the {\em color} of the TO (as $\delta(B-V)$; see Stetson, Vandenberg \& Bolte
1996 for discussion and references). Saviane, Rosenberg \& Piotto (1997)
recently presented a variant of the classical horizontal parameter for the
(V;V-I) plane, i.e. $\delta(V-I)_{2.2}$ the difference in color between the TO
and the RGB at a level 2.2 mag brighter than the TO. The method is very
promising\footnote{However, as all the horizontal methods, it requires an
exceptional accuracy in the determination of the observable, since differences 
of few hundredths of mag in $\delta(V-I)_{2.2}$ can correspond to difference 
in age $>2$ Gyrs.}
since the preliminary results of Saviane, Rosenberg \& Piotto (1997)
suggest that $\delta(V-I)_{2.2}$ is virtually independent of metallicity. 

From our ridge line we obtain $(V-I)_{TO}=0.74\pm 0.02$ and  
$(V-I)_{2.2}=1.07\pm 0.02$ and finally $\delta(V-I)_{2.2}=0.33\pm 0.03$, 
the same as the well known young globular Pal 12. From Fig. 8 of Saviane, 
Rosenberg \& Piotto (1997) and the calibration provided by the same authors we
derived an age difference of $\sim 4 Gyr$ between Sgr and the bulk of the
Galactic globulars, independently of the assumed metallicity.
 
\subsection{Is Pop A a single-age population?}

All the above age determinations are based on the following implicit hypothesis:

\begin{enumerate}
 
\item Pop A is a single-metallicity population or, at least, a meaningful
{\em average metallicity} can be adopted. 

\item Pop A is a single-age population or, at least, a meaningful 
{\em average age} can be derived from the mean TO luminosity or color, once the
metallicity is fixed.

\end{enumerate}

Both of them are in large part false. Pop A cannot be
treated as a single-metallicity population since there are clear
observational evidences that there is a mixture of stars with significantly 
different abundances (Sect. 5).  
The derivation of an {\em average age} from the observed TO can be
completely meaningless, since populations of different age and metallicity can
have the TO placed at the same position.

Given the observed metallicity spread and TO morphology, two possible extreme 
scenarios may be invoked to describe the SFH of Sgr Pop A: 
(1) stars formed $\sim 10 ~Gyr$ ago in a short lapse of time
($\le 1 Gyr$) from a very inhomogeneously enriched interstellar medium, with a
wide range of abundances; (2) stars began to form at very early epoch and 
star formation continued for a quite long  
period ($\ge 4 Gyr$) from a chemically homogeneous interstellar 
medium (ISM) which was progressively enriched.

Option 1 seems contrary to the common wisdom on galaxy evolution, since it is
expected that metallicity would increase with time. Furthermore it would be 
in contrast with the clear Age-Metallicity Relation defined by its globulars 
(MoAL). 
  
\subsubsection{Theoretical expectations}

In Fig. 8 the two scenarios described above are illustrated according to the
expectations of the stellar evolution theory. Here we adopt the
homogeneous set of isochrones by Bertelli et al. (1994), but the results do not
change if other isochrones are used\footnote{Very similar results
have been obtained with the Cassisi et al. (1998) and the Vandenberg
isochrones.}.  

In panel (a) the {\em single-age
+ multi-metallicity} case is presented reporting three isochrones of $age=10
~Gyr$ and with $[Fe/H]=-1.7, ~-1.3, ~-0.7$, respectively.

In panel (b) the {\em multi-age
+ multi-metallicity} case is presented reporting three isochrones of $age=16,
~12, ~8 ~Gyr$ and with $[Fe/H]=-1.7, ~-1.3, ~-0.7$, respectively. 
Younger isochrones has been associated with higher metal content.

The zoomed diagrams provide closer views of the TO region.

\begin{figure}
 \vspace{20pt}
\epsffile{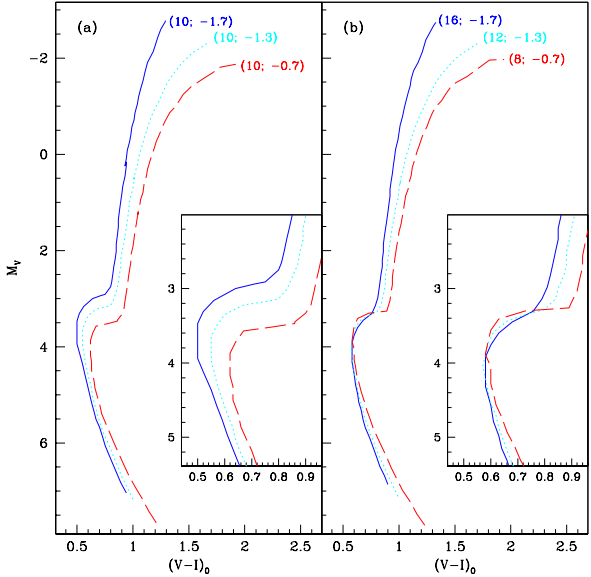}
 \caption{Theoretical expectations for the single-age case (panel a) and for the
 multi-age case (panel b). All isochrones are from the homogeneous set of 
 Bertelli et al. (1994). Each isochrone is labeled according to the notation 
 (age[Gyr]; [Fe/H]). The zoomed diagrams provide a closer view of the TO
 region.}
\end{figure}

It is evident that, while the two groups of isochrones are very
similar in their RGB characteristics, they significantly differ around the TO
level. 

The SGB and Upper MS of the in panel (a) are
clearly detached one from another, with large color spread at the TO level 
($\Delta (V-I)_0 =0.15$), while the isochrones reported in panel (b) are
virtually superposed up to the base of the RGB. This is a well known effect 
(Sandage 1990 and references therein): at fixed age more metal rich populations
have fainter TO points\footnote{ This is one of the aspects of the 
age/metallicity degeneracy, an ubiquitous problem in the study of stellar 
populations (see Renzini \& Fusi Pecci 1988, and references therein).} 

In the SGR34 CMD, the observed color dispersion around the ridge line of the 
stars with $4.0<M_V<3.6$ and $0.0<V-I<1.1$ is $\sigma_{(V-I)}=0.12$, 
nearly equal to the photometric error at this level. Thus the
data are perfectly consistent with nearly null intrinsic width and the extreme 
case {\em single age + multi metallicity} can be probably excluded.
However such relatively high photometric errors at the TO level do not allow a 
clear-cut discrimination between any intermediate occurrence between the 
two extreme cases illustrated in Fig. 8.

\subsubsection{Empirical age dating}

In Fig. 9 the ridge line of SGR34 Pop A (open circles with central dot) is
compared to the template ridge lines (continuous lines; dashed lines are
adopted when the upper part of the line is implemented from a different source,
see Sect. 5). The diameter of the open circles approximately corresponds to the
extent of the uncertainty in the Sgr distance modulus ($\sim \pm 0.3 ~dex$).
In the following, the comparison with the different templates will be performed
discussing the various panels of Fig. 9.

\begin{figure*}
 \vspace{20pt}
\epsffile{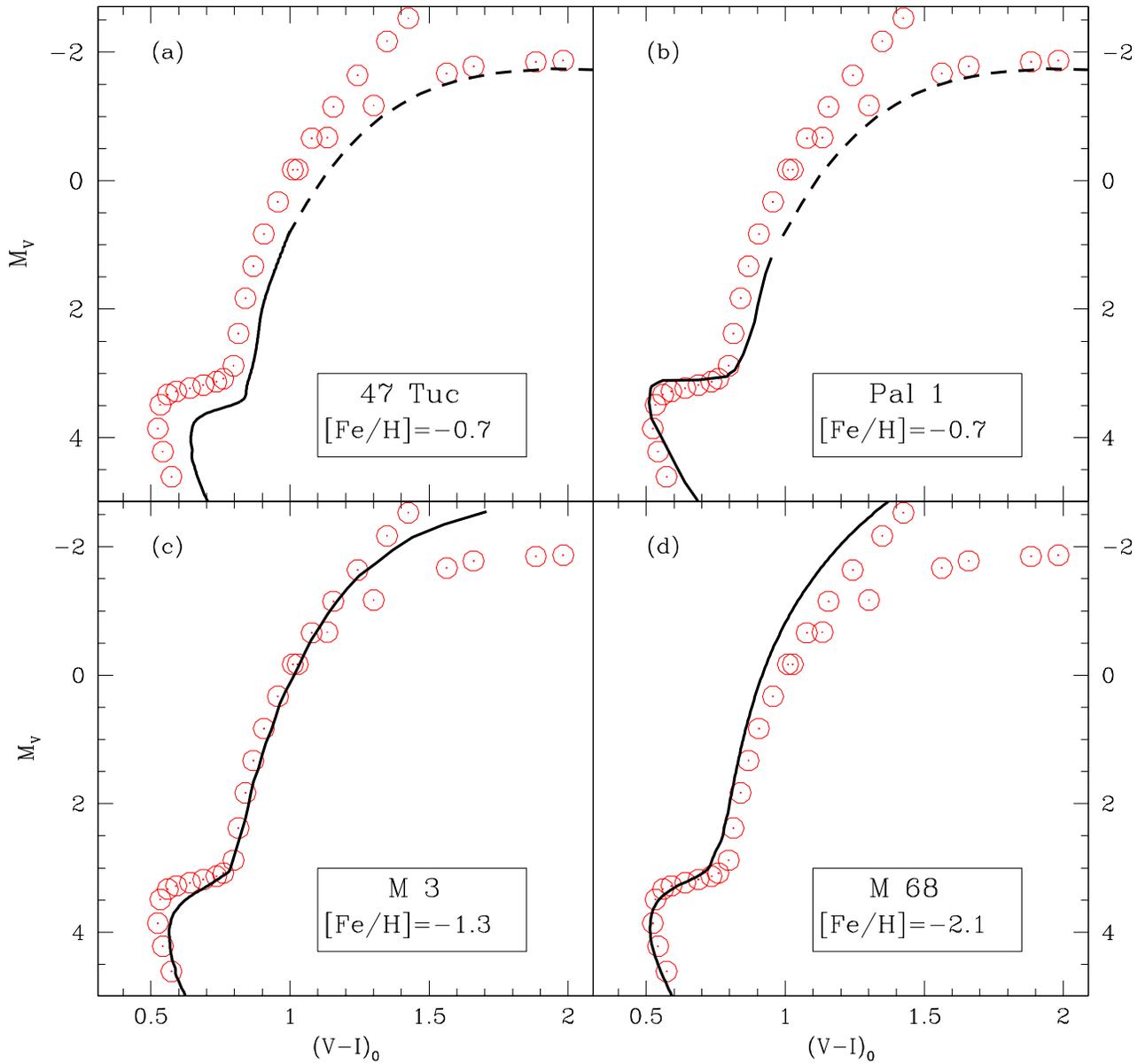}
 \caption{Comparisons between the SGR34 ridge line (open circles with central
 dot) and the ridge lines of the template clusters (continuous lines; the
 dashed line means that part of the ridge line has been complemented with data
 from a different source - see Sect. 5). 
 All the lines have been reported to the
 absolute ($M_V;(V-I)_0$) plane with the assumptions described in Sect. 5.
 The radius of the open circles has approximately the dimension of the error box 
 of the distance modulus of the Sgr galaxy. In the enclosed box in the lower
 left corner of each panel report the name and the metallicity of the template
 (panel (a): 47 Tuc; panel (b): Pal 1; panel (c): M 3; panel (d): M 68).}
\end{figure*}

\begin{itemize}

\item Panel (a), template: 47 Tuc. While this cluster provides a reasonable fit
to the red edge of the RGB, the SGB and UMS sequences are clearly fainter
and redder than those of SGR34. Note that the discrepancy cannot be recovered by
different assumptions on the distance moduli and can be partially recovered 
only by assuming an error in $E(V-I)$ of at least $0.1 ~dex$, that is {\em very
unlikely}. So, it can be safely concluded that {\em the Sgr stars at the same
metallicity of 47 Tuc are significantly younger than this cluster}.

\item Panel (b), template: Pal 1. This cluster has the same metallicity as 
47 Tuc but is $\sim 4 ~Gyr$ younger, and provide a reasonable fit to the SGR34
faint sequences (it appears slightly ``younger'' but the two ridge lines are 
fully compatible within errors). {\em The Sgr stars 
at the same metallicity of 47 Tuc have an age similar to the globular cluster 
Pal 1, i.e. $\sim 4 ~Gyr$ younger than 47 Tuc}.

\item Panel (c), template: M3. This template provides the best fit to the
average RGB ridge line of SGR34, and indeed the metallicity distribution of
Pop A peaks around the metallicity of M3. The fit to the SGB and UMS is not
completely unconsistent but a younger template is probably needed. 
Unfortunately, a safe ``younger'' template at that metallicity is
not available (at least in the V,I passbands).
We tentatively conclude that {\em the stars
of Sgr Pop A with metallicity around $[Fe/H]=-1.3$ are probably
younger than the M3 globular cluster (a classical old one)}. However the age
difference between this component and M3 is probably smaller than the one 
between the metal rich component and 47 Tuc.

\item Panel (d), template: M68. As shown in Sect. 5, the template provides a
good fit to the blue edge of the Pop A RGB and the fit to the TO region is
nearly perfect. We conclude that {\em the most metal poor stars of Sgr Pop
A are as old as the oldest galactic globulars}. Note that also the most metal 
poor globulars belonging to the Sgr galaxy (M54 and Ter 8) are at least as old
as M68 (MoAL, MAL, LS97).

\end{itemize}

The above comparisons are clearly consistent with a scenario of chemical
enrichment during a star formation phase lasted many billion years, in good 
agreement with the Globular Cluster AMR by MoAL. In the
following sections we will couple these results with the metallicity
distribution derived in Sect. 5 to provide possible models of the Star
Formation History of Pop A. 

\subsection{Setting the SF time scales}

Though the SFR had a sharp decrease after the generation of stars at
$[Fe/H]=-0.7/-0.5$, more recent low rate star formation could also take
place in the Sgr galaxy (Pop B). Since this younger population is a minor
component (see PAP-I) it is impossible to state if there was a temporal hiatus 
between the completion of Pop A and the onset of Pop B or if star formation 
continued at a lower rate.
Furthermore, the empirical approach adopted for the analysis of
Pop A cannot be applied to the Blue Plume stars, and we simply try to get an
heuristic view by isochrone fitting. This would provide at least a zero-order
global description of Pop A + Pop B as a whole, giving an idea of the SF
time scales.

In Fig. 10 we have superimposed the isochrones of ($[Fe/H]=-1.7$; age=16 Gyr),
($[Fe/H]=-1.3$; age=12 Gyr) and ($[Fe/H]=-0.7$; age=8 Gyr) from Bertelli et al.
(1994) to the CMD of SGR34. To provide a reasonable fit to all the observed
features we had to shift the isochrones by $\Delta V=-0.1$ and 
$\Delta (V-I)=-0.05$. 

\begin{figure*}
 \vspace{20pt}
\epsffile{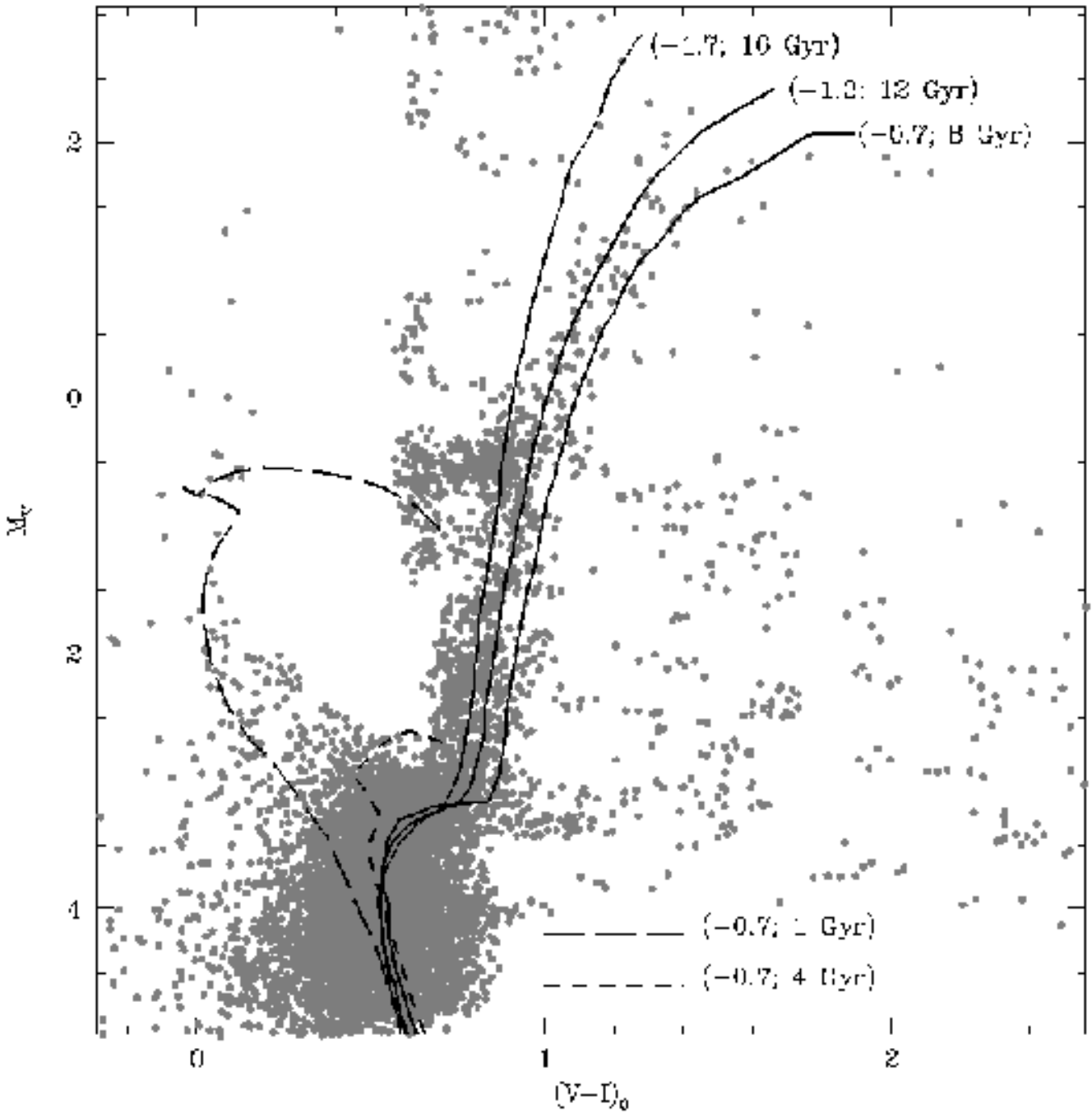}
 \caption{Isochrones from the Bertelli et al. (1994) set superposed to the SGR34
 CMD. The older isochrones (continuous lines) are labeled with the
 corresponding metallicity ([Fe/H]) and age (Gyr). The isochrones bracketing
 the Blue Plume stars have metallicity $[Fe/H]=-0.7$ and ages of 1 Gyr (long
 dashed line) and 4 Gyr (short dashed line), respectively.}
\end{figure*}

Isocrones at $[Fe/H]=-0.7$ of age 1 Gyr (long dashed line) and 4 Gyr 
(short dashed line)
are also reported, and appear to bracket the whole Blue Plume. As we will see
below the adoption of a higher metallicity for Pop B would shift the age
estimates to younger ages.

According to Fig. 10, the main star formation period in the Sgr galaxy could
have lasted for as much as $\sim 8 ~Gyr$ before some circumstance induced a
sharp decrease in the star formation rate. 

In Fig. 11 the CMD of the Blue Plume stars from the SGR34 and SGR12R samples
is shown. This is the most populous CMD of such stars ever obtained.
In the left panel isochrones at $[Fe/H]=-0.7$ and ages of 0.5, 1, 2 and 4 Gyr
(from left to right) are superimposed to the CMD. In the right panel isochrones
of the same ages but with $[Fe/H]=-0.4$ are reported. 

\begin{figure*}
 \vspace{20pt}
\epsffile{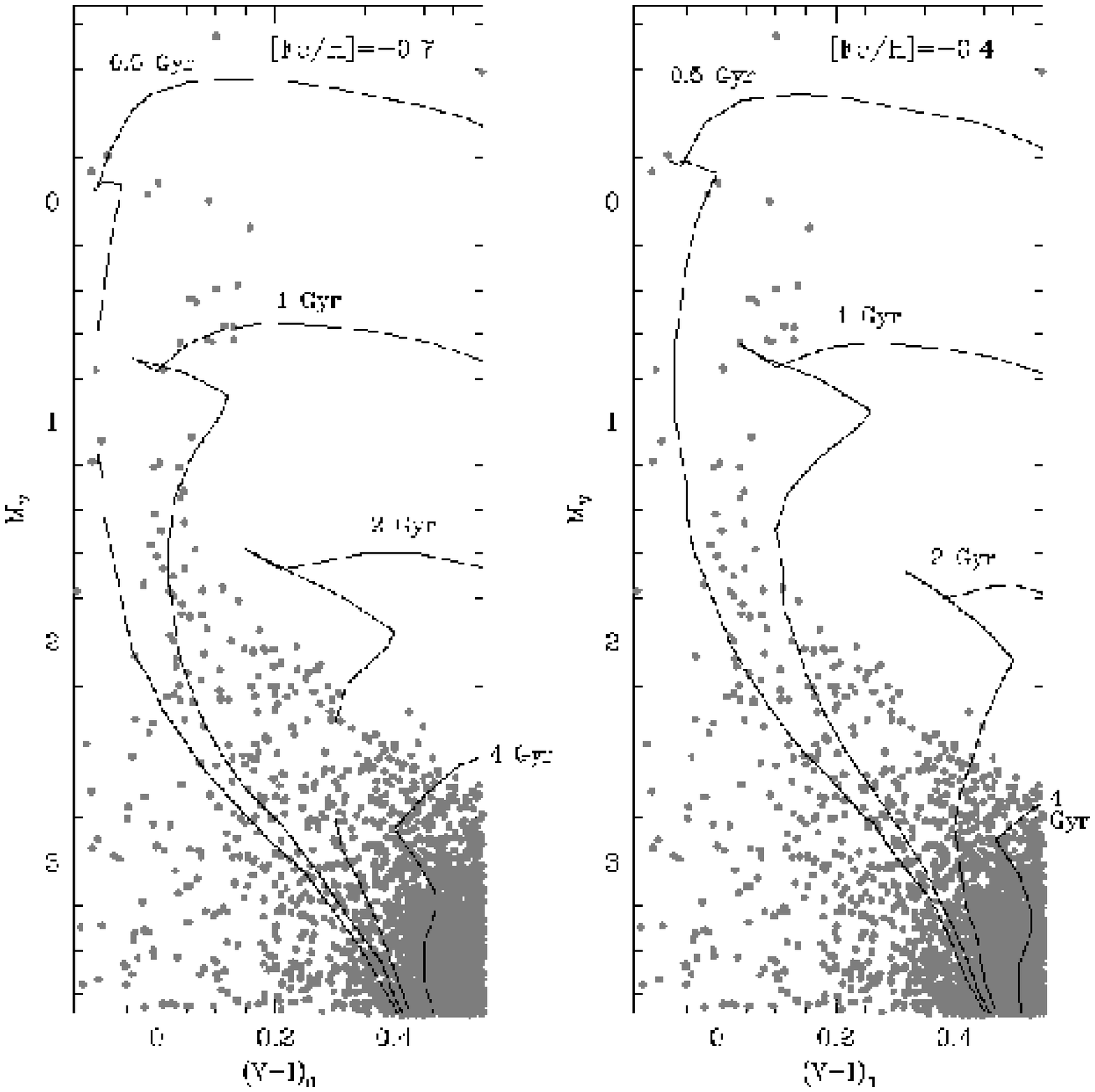}
 \caption{Isochrones from the Bertelli et al. (1994) set superposed to the 
 SGR34+SGR12R CMD, centered on the Blue Plume stars. The isochrones 
 are labeled with the corresponding age (Gyr) and have metallicity 
 $[Fe/H]=-0.7$ (left panel) and $[Fe/H]=-0.4$ (right panel). }
\end{figure*}

From the inspection of Fig. 11 it can be concluded that the observed Blue Plume
is consistent with a remarkably long lasting star formation event ($\sim 3
Gyr$), or with many bursts separated by short lapses of quiescence. 
The time at which star formation definitively stopped depends on the actual 
metallicity of the stars, which is presently unknown. It is worth noting that
if Pop B (or at least part of it) has solar or slightly super-solar metallicity,
as could be argued from the preliminary results by Smecker-Hane et al. (1998;
see also Da Costa 1999), an age as low as 60-100 million years cannot be 
excluded and the duration of the SF episode may shorten to $<0.5 ~gyr$.

\subsection{Star Formation Histories}

Following a suggestion of the Referee, we try to explore the possible Star
Formation Histories of the Sgr Pop A applying the method described in detail by
Rocha-Pinto \& Maciel (1996, 1997). The method is aimed to the recovery of the
stars birthrate as a function of time ($B(t)$) from the presently observed
metallicity distribution and the AMR of the considered stellar system, and it
proved to be remarkably efficient in constraining the SFH of the Galactic Disc 
[see Rocha-Pinto \& Maciel (1996) for assumptions, limitations and tests of the
method].

Three basic elements are required:

\begin{enumerate}

\item A metallicity distribution modeled with a Gaussian function. We fitted
the metallicity distribution of the SGR34+SGR12 fields (taken as representative
of the average Pop A, see Fig. 6) with a Gaussian with peak
$<[Fe/H]>=-1.3$ and $\sigma_{[Fe/H]}=0.45 ~dex$.

\item An Age Metallicity Relation. Given the good agreement between the
SF timescales derived in the previous sections and the GC AMR of MoAL we put
together all these informations to constrain the global AMR. Three
representative points were adopted from the above analysis of the Sgr field
stars, i.e. $([Fe/H]; t_{[gyr]}) = (-2.0; 0), (-1.3; 4), (-0.7; 8)$, and 
four points from the fourth column of Table 5 by MoAL: 
Ter 8 =(-2.0; 0), M54 =(-1.79; 1.5),
Arp 2 =(-1.84; 5) and Ter 7 =(-1.0; 8). In the final time scale $t=0$ occurs 16
gyr ago and the age of Ter 7 and Arp 2 were increased by $0.5 ~gyr$ to impose
exact coevity between Ter 7 and the more metal rich stars of Pop A, in
agreement with the results by MAL. In the following analysis we will try
different simple AMR models wich can reasonably fit the distribution of these 
points in the $t_{[gyr]} ~vs.~ [Fe/H]$ plane.

\item An assumed cosmic scatter in the abundance of the interstellar medium at
any birthtime $\sigma_z$. Some indication on such scatter can be derived from
the typical dispersion of data points in the $t_{[gyr]} ~vs.~ [Fe/H]$ plane.
We adopted two possible options: $\sigma_z=0.2 ~dex$  and $\sigma_z=0.4 ~dex$.     
 
\end{enumerate}

The time resolution of the derived $B(t)$ is strongly limited by the coarseness
and uncertainty of the metallicity distribution and of the AMR, 
and is estimated to be $\ge 2 ~gyr$. 

Fig. 12 reports the results of the experiments. In the left panels the adopted
AMR models (continuous lines) are displayed, the GC points are represented 
by circles, the field points by squares. 
In the right panels the resulting $B(t)$ (normalized to
their peaks) are reproduced: each of them results from the coupling of the
(unique) metallicity distribution with the assumed AMR. The continuous line
represents the $\sigma_z=0.2 ~dex$ option for the cosmic scatter, the dashed
line represents the $\sigma_z=0.4 ~dex$ one. In the upper $x$-axis the
appropriate metallicity scale is schematically reported. It is important to
recall that the latest time reported in the plots ($t=10 ~gyr$) {\em is not the
present time}, which occurs 6 gyr later, in the adopted scale. The plot refers
only to Pop A, whose star formation history ended $\sim 6-8 ~gyr$ ago. 

We shortly discuss the results of Fig. 12, case by case:

Upper couple of panels: the simplest possible model of AMR has been
adopted, a {\bf linear} one. The resulting $B(t)$ show a decline at early and
late times, for both the $\sigma_z$ options. The $\sigma_z=0.2$ case show
a single broad peak at $[Fe/H]\sim -1.3$. The case with large cosmic scatter
produce a double peaked $B(t)$, with $\sim 5 ~gyr$ between the two maxima.
The linear model appears to be an obvious oversemplification of the enrichment
history, resulting in a $B(t)$ function too broad. One has to keep in mind that
in the applied method the AMR states {\em how fast the clock runs at
any given epoch}. The uniform run of the linear clock results in a period of
intense star formation so extended to be incompatible with the observed time
and metallicity scales.

Middle couple of panels: the AMR of the Sgr globulars is optimally fitted
by an {\bf exponential} model. The star formation rate is very low for the
first $\sim 6 ~gyr$, then a relatively short ($\le 2 ~gyr$) and intense period
of star formation occurred with rapid enrichment. This episode ceased rather
abruptly and $B(t)=0$ before $t=9 ~gyr$. The two $\sigma_z$ options give very
similar results, both $B(t)$ peaks at $t\sim 8$ and $[Fe/H]\sim -0.7$. The 
$\sigma_z=0.4$ shows some sign of a second maximum at $[Fe/H]\sim -1.5$.
We regard this scenario as the most likely, since the age and metallicity scale
of the Sgr globulars is much more robustly stated with respect to the field
one.

Lower couple of panels: a {\bf quadratic} AMR is adopted as a best fit
model for the whole data set. Also in this case the star formation rate at
early times is relatively low and the decrease of $B(t)$ after $t=8 ~gyr$ is
rather abrupt. The $\sigma_z=0.4$ case shows again a secondary peak at 
$[Fe/H]\sim -1.5$.

\begin{figure*}
 \vspace{20pt}
\epsffile{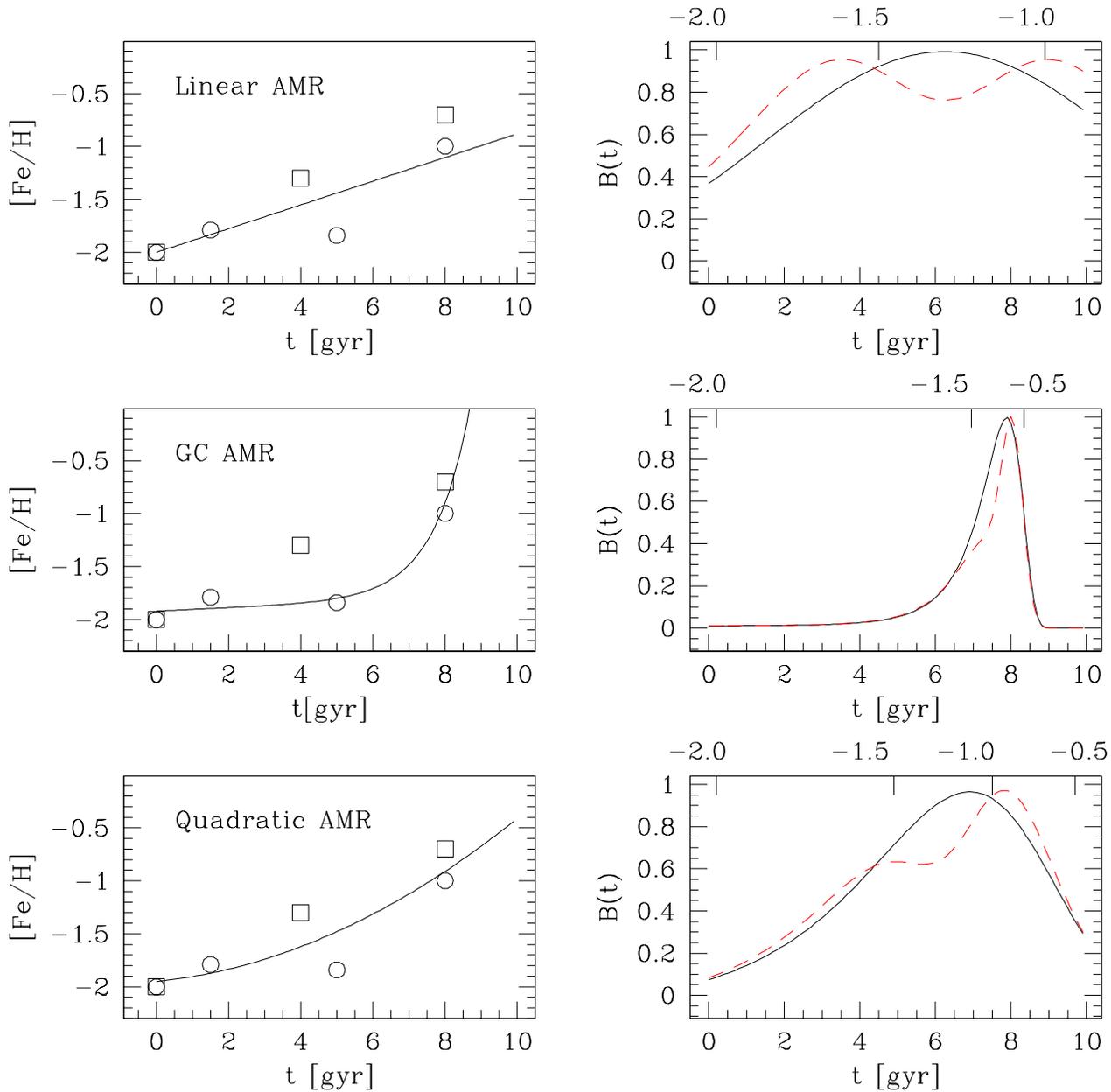}
 \caption{Possible Star Formation Histories for Pop A.
 In the left panels, the adopted AMR models are reported (continuous lines). 
 Squares represent points derived from the present analysis, circles represent 
 the Sgr globular clusters (from MoAL). 
 The corresponding derived SFH are reported in
 the right panels. The continuous line represents the ``low cosmic abundance
 scatter'' case ($\sigma_z=0.2 ~dex$) and the dashed line the ``high cosmic
 abundance scatter'' one ($\sigma_z=0.4 ~dex$).}
\end{figure*}

There are some common characteristics of the derived SFHs which seem to be
rather independent from the assumed AMR:

\begin{itemize}

\item $B(t)$ has a clear main peak, always occurring at $6\le t \le 8 ~gyr$
and $-1.3 \le [Fe/H] \le -0.7$.

\item In any case the SFR at early time is relatively low, and in the two more
realistic cases it rapidly falls after the main maximum.

\item Any eventual secondary peak in the SFR is poorly constrained in time 
(from $t\sim 3$ to $t\sim 7$, depending on the model) but it occurs at
$[Fe/H]\sim -1.5$.

\end{itemize}

In the above scenario the Star Formation event that originated Pop A is
presented as a continuous long lasting episode with a variable SFR, 
but the same general scheme is compatible with many different individual
bursts. The only possibility of drawing a more detailed scheme is to couple
accurate chemical pattern determinations (for instance $[Fe/H]$ and
$\alpha$-elements abundances) with deep photometry for a significant sample of
Sgr Pop A stars.

\section{Summary and Conclusions}

The Color Magnitude Diagrams obtained from the SDGS have been statistically
decontaminated and a detailed analysis of the cleaned diagrams have been
performed, oriented to the search for constraints to the Star Formation History
of the Sagittarius dwarf spheroidal galaxy.

The stellar content of the different regions sampled is remarkably similar and
it can be concluded that star formation in the \sgr has been a very homogeneous
process taking place under the same conditions on spatial scales of the order of
the entire galaxy (see also Mateo, Olszewski \& Morrison 1998).

The main population of the galaxy (Pop A) is composed by stars of metallicity
ranging from $[Fe/H]\sim -2.0$ to $[Fe/H]\sim -0.7$. A stellar component
slightly more metal rich is present in the highest
density clump of the galaxy (SGR12). A coarse metallicity distribution has been
derived. About $60 \%$ of the Pop A stars have $[Fe/H]\ge-1.3$, the metal
poor stars being a relatively minor component of the galaxy.

Based on the large color spread of the RGB (i.e., metallicity spread) and on
the conversely narrow TO and SGB regions (consistent with a single TO point and
SGB sequence, within the observational errors) a significant range in age has
been deduced for Pop A stars, and related with the metallicity range.

The constraints on the AMR have been coupled with the metallicity distribution
to derive the main characteristic of the Star Formation History. A remarkably
robust general scheme emerges. 
 
{\bf Star Formation began in Sgr at low rate, at the epoch of
formation of the oldest globular clusters of the Local Group} (see MoAL,
Buonanno et al. 1998, Stetson et al. 1998). 
{\bf The ISM was progressively enriched
and a main peak of the SFR occurred from 8 to 10 gyr ago when the mean
metallicity was $-1.3 \le [Fe/H] \le -0.7$. During this main episode the
globular cluster Ter 7 was also formed. Then the star formation ceased on
timescales of order $\le 2 ~gyr$ or less, depending on the adopted model.
After this abrupt stop only minor star
formation events took place (still on large spatial scales, see also PAP-I) up 
to 0.5 - 1 Gyr ago (or less) when the gas reservoir of the galaxy was completely 
exhausted} (see Koribalski, Johnston \& Otrupceck 1995).

There are intrinsical difficulties in the study of this particular galaxy,
mainly related to observational problems (Sect. 1) but also due to the peculiar
SFH (Sect. 5 and 6). This is the reason why, despite of the huge efforts 
(SL95, MUSKKK, MAL, Mateo et al. 1996, FAL, IWGIS, LS97), only the {\em average
properties} of the Sgr stellar populations have been derived until now.

Here we have pushed the analysis a significant step beyond, attempting for the
very first time to provide a {\em differential} description of the stellar
content of the \sgr, {\em fitting all its observed properties simultaneously}.
While still coarse and somehow qualitative, this scenario set the 
fundamental framework for the {\em second generation} studies of this
stellar system, switching from a ``monodimensional'' average description
to a ``two-dimensional'' view, from mean age and metallicity to Age-Metallicity
relation and Star Formation History.

This is a key passage if we hope to take full advantage  of the huge
potentiality of the Sgr galaxy as a testbed for our models of dwarf galaxy
formation and evolution and, above all, for our understanding of the links
between star formation and galaxy-galaxy interactions.

\subsection{Are Star Formation History and ``orbital'' history of Sgr coupled?}

Based on radial velocity, a proper motion measurement and few reasonable
assumptions, IWGIS found that the Sgr dSph lies in a very short period orbit 
($P=0.76 ~gyr$) with a perigalactic point located at $\sim 15 ~kpc$ from the
center of the Milky Way. The same authors argues that the orbit has remained
the same for the whole lifetime of the galaxy. On these basis they suggest that
Sgr managed to survive to so many close encounters with the Galaxy thanks to
its high content of dark matter [see also Ibata \& Lewis (1998)]. 
Even if a massive dark matter halo is actually embedding the Sgr stars and
allowed the system to survive such unfavorable conditions, 
the damages of the tidal interactions are evident at the present time 
[see PAP-I and Mateo, Olszewski \& Morrison (1998)] and the possibility that 
Sgr could be only an unbound remnant has been seriously considered 
[Helmi \& White (1999)].  

Now we know that the first stars formed in the Sgr galaxy nearly at the same
time when first Galactic globulars formed. It is very difficult to conceive
that $\sim 15$ close passages to the center of the Milky Way and $\sim
30$ Galactic disc crossings had no effect on the Star Formation and on the 
destiny of the ISM of the Sgr galaxy. 
On the other hand the stellar content and the SFH of
Sgr appears quite ``normal'', i.e. very similar to other dSphs which are
supposed to evolve as unperturbed systems [as Draco or Sculptor, for instance;
see Mateo (1998)]. In particular it is noteworthy that the formation of Pop A
lasted $\sim 8 ~gyr$ with increasing SFR: this means that Sgr preserved 
(and self-enriched) most of its gas content for some 10 orbits. 
To ``preserve its individuality'', this gas must
have efficiently counteracted not only the tidal stress that ultimately
stretched the whole galaxy, but also the ram pressure from the Galactic disc.       

If Sgr evolved unperturbed in the orbit derived by IWGIS then many of our
current believings about the connection between galaxy interactions and star
formation probably should be revised [Schweizer (1998), Kennicut (1998)].

Otherwise, some scenario should be envisaged in which the present orbit is only
the final stage of a significant orbit evolution:

\begin{itemize}

\item Sgr passed most of its life in a wider orbit since some
event [as for example a close encounter with the Magellanic Clouds, as
suggested by Zhao (1998)] induced a significant orbital decay. One can concoct
that the first close passage to the Galactic center coincided with the end of
the star formation period that formed Pop A, or with the recent gasp which
originated Pop B. The scenario proposed by Zhao (1998) could be consistent with
the second hypothesis.

\item Sgr originated from a gas cloud wich is completing just now its process
of collapse toward the potential well of the Galaxy. The collapse time is
function of the distance from the ``center of the collapse''. If at the
``start'' of the Galactic collapse the proto-Sgr
found itself in the remote outskirts of the mass distribution that formed the
Milky Way, it could have been left behind by the bulk of the proto-Galaxy that
collapsed much more rapidly: in a free fall collapse regime 
[Eggen, Lynden Bell, Sandage (1962)] the collapse time at 40 kpc from the
center is $t_{ff}\sim 1 ~gyr$ while at 150 kpc is $t_{ff}\sim 10 ~gyr$.
Sgr could have travelled for many gyr - evolving as an unperturbed, isolated 
dSph - before reaching the inner regions of the Galaxy [see Blitz et al (1998), 
for a more general view of similar phaenomena].

The first close encounter with the Milky Way could
have lead to a strong tidal heating of Sgr, which bounded the galaxy on the 
present, low energy orbit. Also in this case, the onset of the SF episode 
associated with Pop B could have been triggered by this recent first interaction
with the Galaxy. 

\end{itemize}

\section*{Acknowledgments}

We are indebted to an anonymous Referee for his contribution in improving the
final version of the manuscript.
We thank Flavio Fusi Pecci, Ken Mighell and, in particular, 
Monica Tosi for many useful discussions, Santi Cassisi and Don A. Vandenberg
for providing their theoretical isochrones, Alfred Rosenberg for providing
many templates and ridge lines in machine readable format.
A very special thank is owed to Livia Origlia for a careful reading of a draft
version of the manuscript.   

Much of the data analysis has been made easier by the computer codes developed
at the {\em Osservatorio Astronomico di Bologna} by Paolo Montegriffo.
This research has made use of NASA's Astrophysics Data System Abstract Service.

This research has been partially funded by a Grant of the {\it Ministero delle 
Universit\`a e della Ricerca Scientifica e Tecnologica} (MURST) assigned to the
project {\em Stellar Evolution} (national coordinator: Prof. V. Castellani).

%


\end{document}